\documentclass[twoside,letterpaper]{vldb}

\makeatletter
\newif\if@restonecol
\makeatother

\usepackage{multirow}
\usepackage{enumerate}
\usepackage{times}
\usepackage[vlined,algoruled]{algorithm2e}
\usepackage{balance}
\SetKwInOut{Input}{Input}
\SetKwInOut{Output}{Output}

\SetKw{Break}{break}

\SetKw{Continue}{continue}

\SetKwFunction{Pairwise}{PairwiseDetection}
\SetKwFunction{Originals}{FindOriginals}
\SetKwFunction{Global}{GlobalPr}
\SetKwFunction{Max}{max}

\SetCommentSty{textsl}
\SetAlgoSkip{1pt}

\newcommand{\eat}[1]{}
\newcommand{\ie}{{\em i.e.}}
\newcommand{\eg}{{\em e.g.}}

\newcommand{\tightlist}{\itemsep=-3pt}

\begin{document}

\title{Truth Finding on the Deep Web: Is the Problem Solved?}

\numberofauthors{3}

\author{
\alignauthor Xian Li\\
       \affaddr{SUNY at Binghamton}\\
       \email{\small xianli@cs.binghamton.edu}
\alignauthor Xin Luna Dong\\
\affaddr{AT\&T Labs-Research}\\
\email{\small lunadong@research.att.com}
\alignauthor Kenneth Lyons\\
\affaddr{AT\&T Labs-Research}\\
\email{\small kbl@research.att.com}
\and\alignauthor Weiyi Meng\\
       \affaddr{SUNY at Binghamton}\\
       \email{\small meng@cs.binghamton.edu}
\alignauthor
Divesh Srivastava\\
       \affaddr{AT\&T Labs-Research}\\
       \email{\small divesh@research.att.com}
}

\maketitle
\begin{abstract}
{\small The amount of useful information available on the Web
has been growing at a dramatic pace in recent years and people
rely more and more on the Web to fulfill their 
information needs. In this paper, we study truthfulness of
Deep Web data in two domains where we believed data are fairly
clean and data quality is important to people's lives: 
{\em Stock} and {\em Flight}. To our surprise, we
observed a large amount of inconsistency on data from different
sources and also some sources with quite low accuracy. We further
applied on these two data sets state-of-the-art 
{\em data fusion} methods that aim at
resolving conflicts and finding the truth,
analyzed their strengths and limitations, and suggested
promising research directions. We wish our study can increase 
awareness of the seriousness of conflicting data on the Web
and in turn inspire more research in our community to tackle
this problem.}
\end{abstract}

\section{Introduction}
\label{sec:intro}
The Web has been changing our lives enormously. 
The amount of useful information available on the Web 
has been growing at a dramatic pace in recent years. 
In a variety of domains, such as science, business, technology,
arts, entertainment, government, sports, and tourism, 
people rely on the Web to fulfill their information needs.
Compared with traditional media, information on the Web can
be published fast, but with fewer guarantees on quality and credibility.
While conflicting information is observed frequently on the Web,
typical users still trust Web data. 
In this paper we try to understand the truthfulness of Web data
and how well existing techniques can resolve conflicts
from multiple Web sources.

This paper focuses on Deep Web data, where data are stored in underlying
databases and queried using Web forms. We considered two domains,
{\em Stock} and {\em Flight}, where we believed data are fairly 
clean because incorrect values can have a big (unpleasant)
effect on people's lives. As we shall show soon, data for these
two domains also show many different features.

We first answer the following questions.
Are the data consistent? Are correct data provided by
the majority of the sources? Are the sources highly accurate?
Is there an authoritative source that we can trust and ignore
all other sources? Are sources sharing data with or copying from 
each other?
 
Our observations are quite surprising. Even for these domains that
most people consider as highly reliable, we observed a large amount 
of inconsistency: for 70\% data items more than one value is provided.
Among them, nearly 50\% are caused by various kinds of ambiguity,
although we have tried our best to resolve heterogeneity over
attributes and instances; 20\% are caused by out-of-date data;
and 30\% seem to be caused purely by mistakes. Only 70\% correct
values are provided by the majority of the sources (over half 
of the sources); and over 10\% of them are not even provided by more 
sources than their alternative values are.
Although well-known authoritative sources, such as {\em Google 
Finance} for stock and {\em Orbitz} for flight, 
often have fairly high accuracy, 
they are not perfect and often do not have full coverage, so
it is hard to recommend one as the ``only'' source that users
need to care about. Meanwhile,  there are many sources with
low and unstable quality. Finally, we did observe
data sharing between sources, and often on low-quality data,
making it even harder to find the truths on the Web.

Recently, many {\em data fusion} techniques have been proposed 
to resolve conflicts and find the truth~\cite{BCM+10, BN08, DBS09a, 
DBS09b, DN09, GAMS10, PR10, PR11, WM07, WM11, YHY08, YT11, ZRHG12}. 
We next investigate how they perform on our data sets and
answer the following questions. Are these techniques effective?
Which technique among the many performs the best? How much
do the best achievable results improve over trusting data
from a single source? Is there a need and is
there space for improvement?

Our investigation shows both strengths and limitations of the
current state-of-the-art fusion techniques. On one hand, 
these techniques perform
quite well in general, finding correct values for 96\% data items
on average. On the other hand, we observed a lot of instability
among the methods and 
we did not find one method that is consistently better than others.
While it appears that considering trustworthiness
of sources, copying or data sharing between sources, similarity
and formatting of data are helpful in improving accuracy, it is essential that accurate information on source trustworthiness and copying between sources is used; otherwise, fusion accuracy can even be harmed. According to our observations, we identify 
the problem areas that need further improvement.
\eat{are promising to help, not being able
to compute the correct source trustworthiness or to discover the correct
copying relationships in some cases can even hurt.}

\smallskip
\noindent
{\bf Related work:} Dalvi et al.~\cite{DMP12} studied redundancy
of structured data on the Web but did not consider the consistency aspect.
Existing works on data fusion (\cite{BN08, DN09} as surveys and 
\cite{GAMS10, PR10, PR11, WM11, YT11, ZRHG12} as recent works)
have experimented on data collected from the Web
in domains such as {\em book, restaurant} and {\em sports}.
Our work is different in three aspects. First, 
we are the first to quantify and study
consistency of Deep Web data. Second, we are the first to 
compare all fusion methods proposed up to date empirically.
Finally, we focus on two domains where we believed data should be
quite clean and correct values are more critical.
We wish our study on these two domains can increase awareness of
the seriousness of conflicting data on the Web and inspire
more research in our community to tackle this problem. 

\eat{
May consider adding some sentences here to state the significance of this work: By reporting our analysis of web data in the two domains, we hope that we can increase awareness of the seriousness of conflicting data on the Web which in turn can inspire more research in our community to tackle this problem. By evaluating and comparing the state-of-the-art fusion algorithms, we aim to identify the problem areas that need further attention. 
}

\smallskip
In the rest of the paper, Section~\ref{sec:collection} describes 
the data we considered, Section~\ref{sec:quality} describes 
our observations on data quality, Section~\ref{sec:fusion}
compares results of various fusion methods, Section~\ref{sec:extension}
discusses future research challenges, and
Section~\ref{sec:conclude} concludes.

\section{Problem Definition and Data Sets}
\label{sec:collection}

We start with defining how we model data from the Deep Web
and describing our data collections\footnote{\small Our data
are available at http://lunadong.com/fusionDataSets.htm.}.

\subsection{Data model}
We consider Deep Web sources in a particular {\em domain},
such as flights.
For each domain, we consider {\em objects} of the same type, 
each corresponding to a real-world entity. For example,
\eat{an object in the book domain can be a book with a particular ISBN;}
an object in the flight domain can be a particular flight on a particular day. 
Each object can be described by a set of {\em attributes}. 
For example, 
\eat{a book can be described by its title, list of 
authors, publisher, publish year, etc;}
a particular flight can be described
by scheduled departure time, actual departure time, etc.
We call a particular attribute of a particular object a 
{\em data item}. We assume that each data item is associated with
a single {\em true value} that reflects the real world.
For example, the true value for
the actual departure time of a flight is the minute that the airplane 
leaves the gate on the specific day.

Each data source can provide a subset of objects in a particular
domain and can provide values of a subset of attributes 
for each object. Data sources have heterogeneity at three levels.
First, at the schema level, they may structure the data differently
and name an attribute differently.
Second, at the instance level, they may represent an object
differently. This is less of a problem for some domains where
each object has a unique ID, such as 
stock ticker symbol, but more of a problem for other domains such as
business listings, where a business is identified by its name,
address, phone number, business category, etc.
Third, at the value level, some of the provided values 
might be exactly the true values, 
some might be very close to (or different
representations of) the true values, but some might be
very different from the true values. 
In this paper, we manually resolve heterogeneity at the schema
level and instance level whenever possible, and focus on heterogeneity at the 
value level, such as variety and correctness of provided values.

\eat{
. At a high level, we study the following features of
the data, whose formal definitions are given later as we describe
the results. 
\begin{itemize}\tightlist
  \item {\em Redundancy}: The percentage of sources that provide
    the same data item.
  \item {\em Variety}: The difference of values provided by different
    sources on the same data item.
  \item {\em Accuracy}: The consistency of the data provided
    by a source or from the integrated data with the ground truth. 
\end{itemize}
}

\begin{table}
\centering
\vspace{-.15in}
{\small
\caption{Overview of data collections
\label{tbl:dataset}}
\begin{tabular}{|c|c|c|c|c|c|c|c|}
\hline
           & \multirow{2}{*}{Srcs} & \multirow{2}{*}{Period}   & \multirow{2}{*}{Objects}  &  Local & Global & Considered \\
           & & & & attrs & attrs & items \\
\hline
Stock   &   55   &  July 2011 & 1000*21 & 333 & 153 & 16000*21\\
Flight   &   38   &  Dec 2011  & 1200*31 & 43 & 15  & 7200*31         \\
\eat{Book    &    ?     &  Nov, 2011 & ?	   &   ?    & ? & ?   \\}
\hline
\end{tabular}}
\vspace{-.15in}
\end{table}
\begin{figure*}[t]
\begin{center}
\begin{minipage}{.32\linewidth}
\begin{center}
%
%
\includegraphics[width=5.5cm]{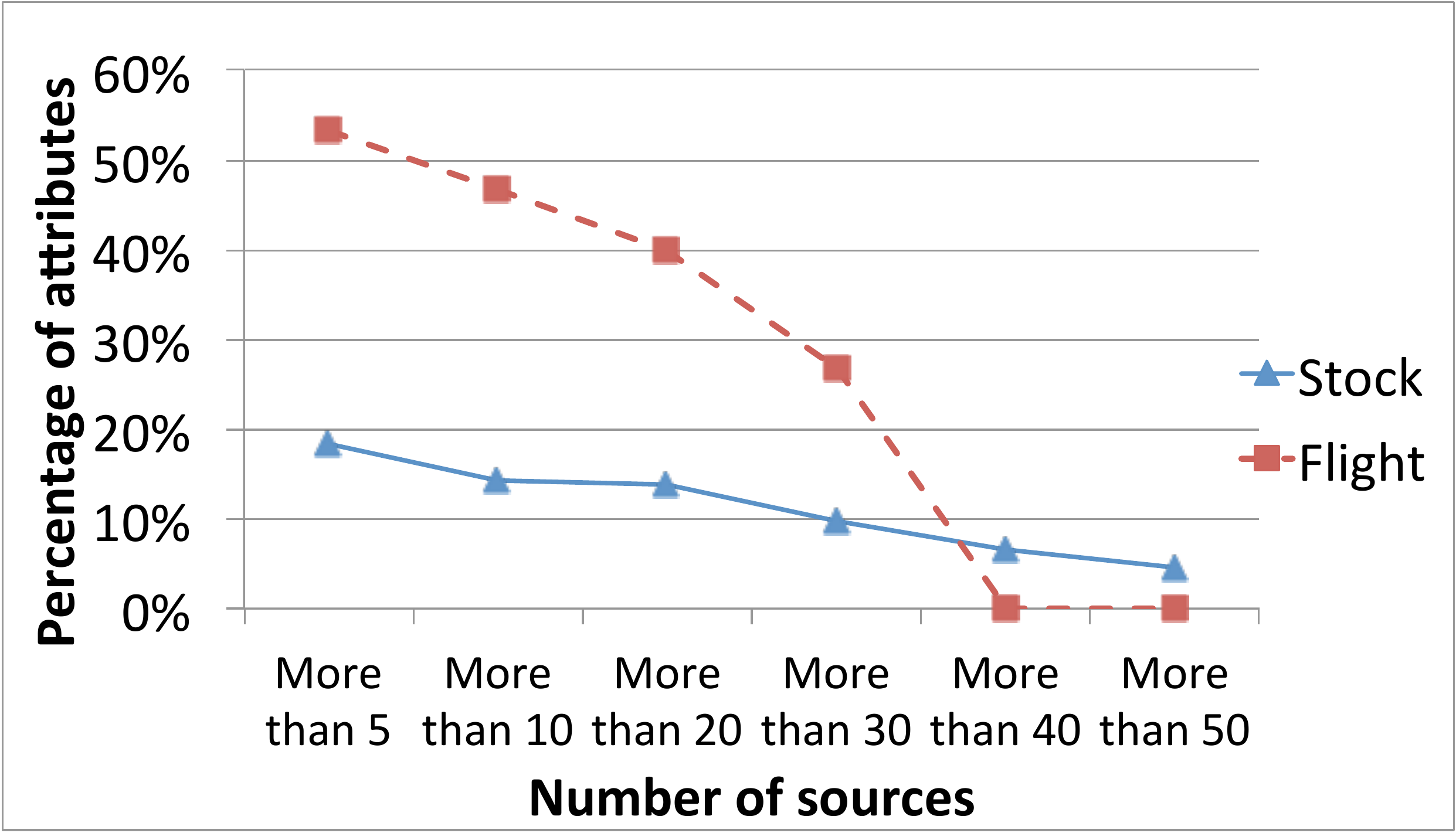}
\vspace{-.1in}
{\caption{\label{fig:covAttr}Attribute coverage.}}
\end{center}
\end{minipage}
\hfill
\begin{minipage}{.32\linewidth}
\begin{center}
%
%
\includegraphics[width=5.5cm]{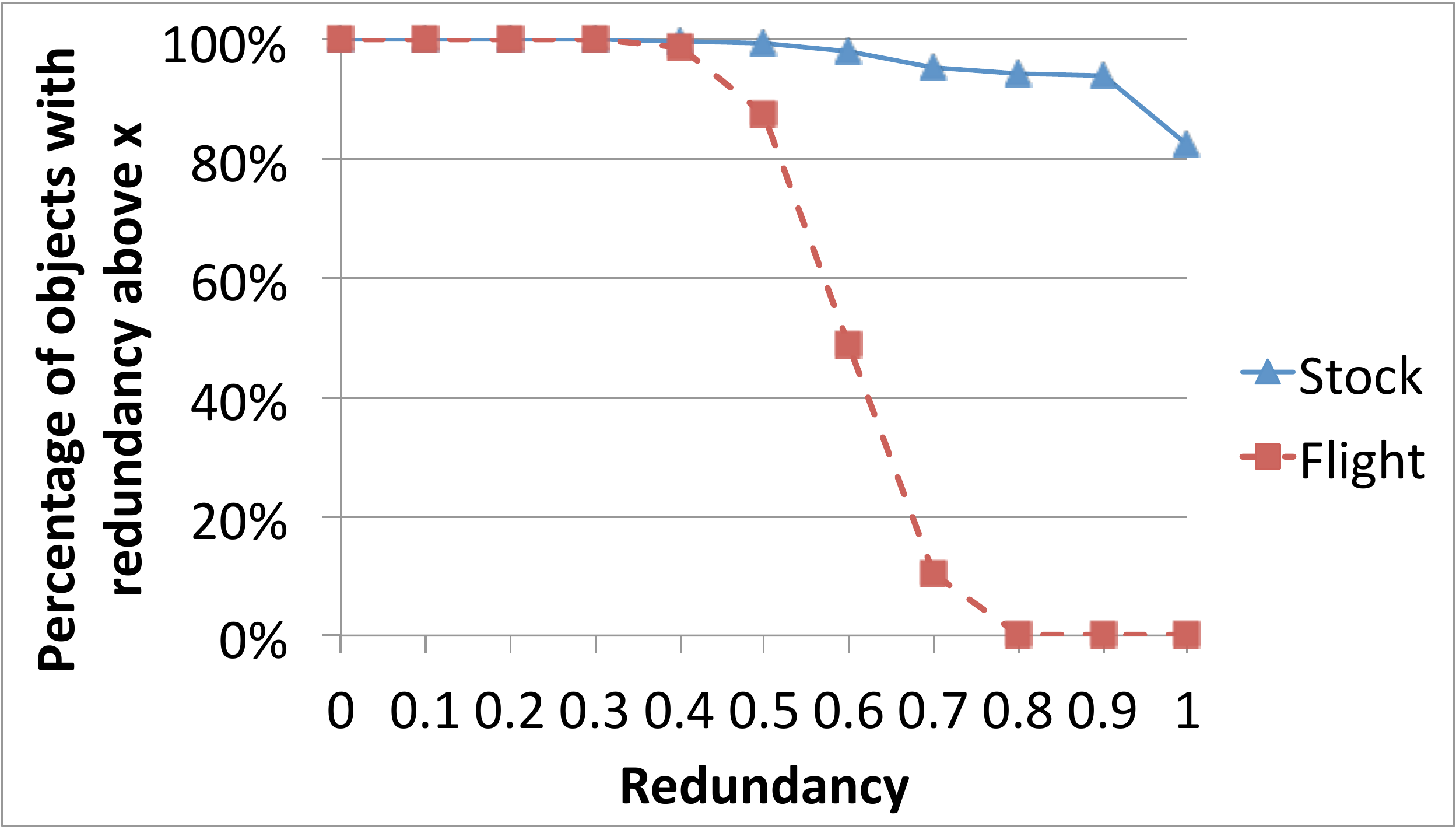}
%
%
\vspace{-.1in}
{\caption{\label{fig:covObj}Object redundancy.}}
\end{center}
\end{minipage}
\hfill
\begin{minipage}{.32\linewidth}
\begin{center}
%
%
\includegraphics[width=5.5cm]{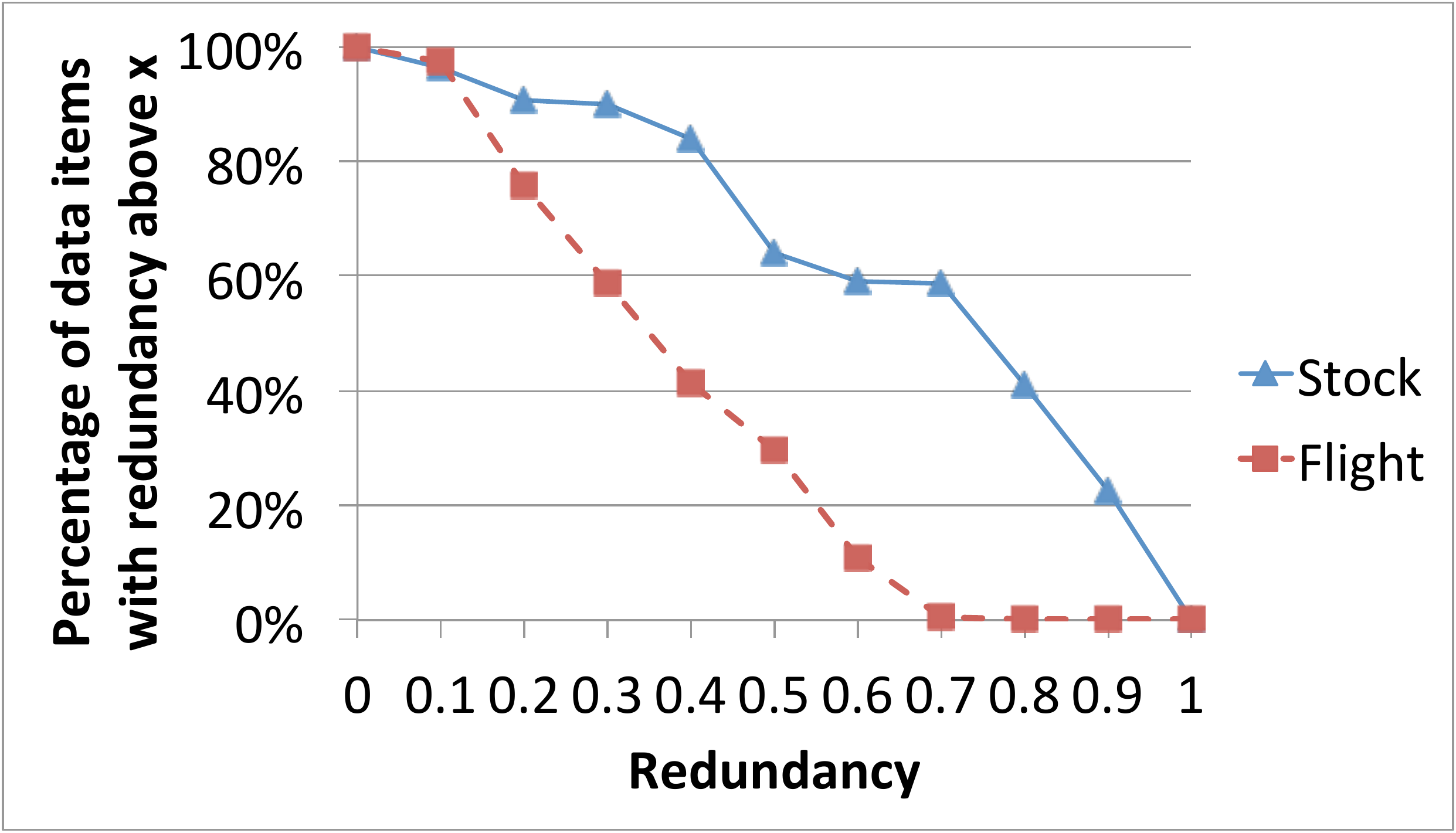}
%
%
\vspace{-.1in}
{\caption{\label{fig:covItem}Data-item redundancy.}}
\end{center}
\end{minipage}
\end{center}
\vspace{-.25in}
\end{figure*}
\eat{
\begin{figure*}[t]
\vspace{-1.0in}
\begin{minipage}[b]{0.3\textwidth}
\includegraphics[width=2.2in]{figs/attribute_stock.pdf}
\vspace{-.25in} 
\end{minipage}
\hfill
\begin{minipage}[b]{0.3\textwidth}
\hspace{-.2in}
\includegraphics[width=2.2in]{figs/attribute_flight.pdf}
\vspace{-.1in} 
\end{minipage}
\hfill
\begin{minipage}[b]{0.3\textwidth}
\hspace{-.2in}
\includegraphics[width=2.2in]{figs/attribute_flight.pdf}
\vspace{-.15in} 
\end{minipage}
\vspace{-1in}
\caption {Attribute Label Distribution}
\label{fig:attribute}
\end{figure*}
}
\subsection{Data collections}
\eat{
We studied two collections of data, whose features are shown 
in Table~\ref{tbl:dataset}. 
We have two criteria in selecting
the data sets. First, we wish to cover different types of domains.
Second, we wish that each object in the data collection can be
easily identified by some ID, such as stock ticker symbol.
}

We consider two data collections from 
{\em stock} and {\em flight} domains where we believed data are fairly
clean and we deem data quality very important. 
Table~\ref{tbl:dataset} shows some statistics of the data.

\smallskip
\noindent
{\em \bf Stock data:}
The first data set contains 55 sources in the {\em Stock} domain.
We chose these sources as follows. We searched ``stock price
quotes'' and ``AAPL quotes'' on {\em Google} and {\em Yahoo},
and collected the deep-web sources from the top 200 returned results.
There were 89 such sources in total. Among them, 76 use the {\tt GET}
method (\ie, the form data are encoded in the URL) and 13 use
the {\tt POST} method (\ie, the form data appear in a
message body). We focused on the former 76 sources, for which data
extraction poses fewer problems. 
Among them, 17 use Javascript to dynamically
generate data and 4 rejected our crawling queries. So we focused on the remaining
 55 sources. 
These sources include some popular financial aggregators such as
{\em Yahoo! Finance}, {\em Google Finance}, and {\em MSN Money},
official stock-market websites such as {\em NASDAQ}, 
and financial-news websites such as {\em Bloomberg} and {\em MarketWatch}. 

We focused on 1000 stocks, including the 30 symbols from Dow Jones Index, 
the 100 symbols from NASDAQ Index (3 symbols appear in both Dow Jones
and NASDAQ), and randomly chosen 873 symbols from the other 
symbols in Russell 3000. Every weekday in July 2011 we searched each
stock symbol on each data source, downloaded the returned web pages,
and parsed the DOM trees to extract the attribute-value pairs. 
We collected data one hour after the stock 
market closes on each day to minimize the difference caused by 
different crawling times. Thus, each object is a particular stock
on a particular day.

We observe very different attributes from different sources about
the stocks: the number of attributes provided by a source ranges
from 3 to 71, and there are in total 333 attributes. Some of the
attributes have the same semantics but are named differently.
After we matched them manually, there are 153 attributes.
We call attributes before the manual matching {\em local attributes}
and those after the matching {\em global attributes}.
Figure~\ref{fig:covAttr} shows 
the number of providers for each global attribute.
The distribution observes {\em Zipf's law};
that is, only a small portion of attributes 
have a high coverage and most of the ``tail'' attributes 
have a low coverage. In fact, 21 attributes (13.7\%) are provided by 
at least one third of the sources and over 86\% are provided by 
less than 25\% of the sources. Among the 21 attributes, the values of
5 attributes can keep changing after market close due to after-hours 
trading. In our analysis we focus on the remaining 16 attributes,
listed in Table~\ref{tbl:stockSymbol}.
For each attribute, we normalized values to the same format
(\eg, ``6.7M'', ``6,700,000'', and ``6700000'' are considered as the same value).

\begin{table}
\vspace{-.15in}
\centering
{\small
\caption{Examined attributes for {\em Stock}.
\label{tbl:stockSymbol}}
\begin{tabular}{|c|c|c|c|}
\hline
Last price& Open price & Today's change (\%)  &     Today's change(\$)\\
Market cap   &  Volume   &  Today's high price & Today's low price \\
Dividend   &  Yield   &  52-week high price  & 52-week low price \\
EPS    &    P/E     &  Shares outstanding & Previous close \\
\hline
\end{tabular}}
\vspace{-.15in}
\end{table}
For purposes of evaluation we generated a gold standard
for the 100 NASDAQ symbols and another 100 randomly selected symbols.
We took the voting results
from 5 popular financial websites: {\em NASDAQ},
{\em Yahoo! Finance}, {\em Google Finance}, {\em MSN Money},
and {\em Bloomberg}; we voted
only on data items provided by at least three sources.
The values in the gold standard are also normalized. 

\smallskip
\noindent
{\em \bf Flight data:}
The second data set contains 38 sources from the flight domain. 
We chose the sources in a similar way as in the stock domain
and the keyword query we used is ``flight status''.
The sources we selected include 3 airline websites 
({\em AA, UA, Continental}), 8 airport websites (such as {\em SFO, DEN}), 
and 27 third-party websites, including {\em Orbitz, Travelocity},
etc. 

We focused on 1200 flights departing from or arriving at the 
{\em hub} airports of the three airlines ({\em AA, UA,} and {\em Continental}). 
We grouped the flights into batches according to their scheduled
arrival time,
collected data for each batch one hour after the latest scheduled
arrival time every day in Dec 2011.
Thus, each object is a particular flight on a particular day. 
We extracted data and normalized the values in the same way
as in the {\em Stock} domain.

There are 43 local attributes and 15 global attributes 
(distribution shown in Figure~\ref{fig:covAttr}).
Each source covers 4 to 15 attributes. The distribution of the
attributes also observes {\em Zipf's law}: 
6 global attributes (40\%) are
provided by more than half of the sources while 53\% of
the attributes are provided by less than 25\% sources.
We focus on the 6 popular attributes in our analysis,
including {\em scheduled departure/arrival time, actual departure/arrival
time}, and {\em departure/arrival gate}. 
We took the data provided by the three airline websites 
on 100 randomly selected flights as the gold standard.

\smallskip
\noindent 
{\bf Summary and comparison:} 
\eat{Among the three data sets, stock data 
and flight data fall in the changing-value domain, while the book
information part of the book data falls in the fixed-value domain
and the sale information part of the book data falls in the 
evolving-value domain (recall that some information, such as 
price of a book, is not updated constantly from each website).
In each collection, objects can be clearly distinguished from each other.
}
In both data collections objects are easily distinguishable from 
each other: a stock object can be identified by date and stock symbol,
and a flight object can be identified by date, flight number, 
and departure city (different flights departing from different cities
may have the same flight number). On the other hand,
we observe a lot of heterogeneity for attributes and value formatting;
we have tried our best to resolve the heterogeneity manually.
In both domains we observe that the distributions of the
attributes observe {\em Zipf's Law}
and only a small percentage of attributes are popular among all sources.
The {\em Stock} data set is larger than the {\em Flight} 
data set with respect to both the number of sources and 
the number of data items we consider.

Note that generating gold standards is challenging when we cannot
observe the real world in person but
have to trust some particular sources.
Since every source can make mistakes, we do voting on authority
sources when appropriate.

\section{Web Data Quality}	
\label{sec:quality}
We first ask ourselves the following four questions about Deep Web data
and answer them in this section.
\begin{enumerate}\tightlist
  \item {\em Are there a lot of redundant data on the Web?} In other words,
    are there many different sources providing data on the same data item?
  \item {\em Are the data consistent?} In other words, are the data provided
    by different sources on the same data item the same
    and if not, are the values provided by the majority of the sources 
    the true values?
  \item {\em Does each source provide data of high quality in terms of correctness
    and is the quality consistent over time?}
    In other words, how consistent are the data of a source compared 
    with a gold standard? And how does this change over time?
  \item {\em Is there any copying?} In other words, is there any 
    copying among the sources and if we remove them, are the majority
    values from the remaining sources true?
\end{enumerate}

We report detailed results on a
randomly chosen data set for each domain: the data of 7/7/2011
for {\em Stock} and the data of 12/8/2011 for {\em Flight}. In addition, 
we report the trend on all collected data (collected on different days).


\subsection{Data redundancy}
\label{sec:redundancy}
We first examine redundancy of the data. The {\em object (resp., data-item) 
redundancy} is defined as the percentage of sources that provide a particular 
object (resp., data item). Figure~\ref{fig:covObj} and 
Figure~\ref{fig:covItem} show the redundancy
on the objects and data items that we examined;
note that the overall redundancy can be much lower. 

For the {\em Stock} domain, we observe a very high redundancy at the 
object level: about 16\% of the sources provide
all 1000 stocks and all sources provide over 90\% of the stocks;
on the other hand, almost all stocks have a redundancy over 50\%,
and 83\% of the stocks have a full redundancy (\ie, provided by all sources).
The redundancy at the data-item level is much lower
because different sources can provide different sets of attributes.
We observe that 80\% of the sources cover over half of the data items,
while 64\% of the data items have a redundancy of over 50\%.

For the {\em Flight} domain, we observe a lower redundancy.
At the object level, 36\% of the sources cover 90\% of the flights
and 60\% of the sources cover more than half of the flights;
on the other hand, 87\% of the flights have a redundancy
of over 50\%, and each flight has a redundancy of over 30\%.
At the data-item level, only 28\% of the sources provide more than
half of the data items, and only 29\% of the data items have 
a redundancy of over 50\%. This low redundancy is because 
an airline or airport web site provides information 
only on flights related to the particular airline or airport.

\smallskip
\noindent
{\bf Summary and comparison:} Overall we observe a large redundancy
over various domains: on average each data item has a redundancy 
of 66\% for {\em Stock} and 32\% for {\em Flight}.
The redundancy neither is uniform across different data items,
nor observes {\em Zipf's Law}: very small portions of 
data items have very high redundancy, very small portions have 
very low redundancy, and most fall in between (for different
domains, ``high'' and ``low'' can mean slightly different numbers). 

\subsection{Data consistency}
\label{sec:conflicts}
We next examine consistency of the data. We start with measuring inconsistency
of the values provided on each data item and consider the following
three measures. Specifically, we consider data item $d$
and we denote by $\bar V(d)$ the set of values provided
by various sources on $d$.
\begin{itemize}\tightlist
  \item {\em Number of values:} We report the number of different 
    values provided on $d$; that is, we report $|\bar V(d)|$,
    the size of $\bar V(d)$.
  \item {\em Entropy:} 
    We quantify the {\em distribution} of the various values  
    by {\em entropy}~\cite{SV09}; intuitively, the higher the
    inconsistency, the higher the entropy. 
    If we denote by $\bar S(d)$ the set of sources that provide 
    item $d$, and by $\bar S(d,v)$ the set of sources that 
    provide value $v$ on $d$, we compute the entropy on $d$ as 

\vspace{-.15in}
{\small
\begin{eqnarray}
E(d)=-\sum_{v \in \bar V(d)}{|\bar S(d,v)| \over |\bar S(d)|}\log{|\bar S(d,v)| \over |\bar S(d)|}.
\label{eqn:entropy}
\end{eqnarray}
}
\vspace{-.15in}
\eat{
    Entropy is always non-negative and it 
    combines the number of values and the distribution
    of the values: if there is a single value provided for $d$, 
    intuitively we have the lowest inconsistency and the entropy is 0;
    for different values with uniform distribution, 
    the larger the number of values, intuitively the higher the inconsistency
    and the higher the entropy;
    for the same number of values, the more uniformly distributed
    are the values, intuitively the higher the inconsistency and actually
    the higher the entropy. }
  \item {\em Deviation:} 
    For data items with conflicting numerical values 
    we additionally measure the
    difference of the values by deviation.
    Among different values for $d$,
    we choose the {\em dominant value} $v_0$ as the one
    with the largest number of providers; that is,
    $v_0=\arg\max_{v \in \bar V(d)}|\bar S(d,v)|$. 
    We compute the deviation for $d$ as the relative deviation w.r.t. $v_0$:

\vspace{-.15in}
{\small
\begin{eqnarray}
D(d)=\sqrt{{1 \over |\bar V(d)|}\sum_{v \in \bar V(d)}({v-v_0 \over v_0})^2}.
\label{eqn:deviation}
\end{eqnarray}
}
\vspace{-.15in}

    We measure deviation for time similarly but use absolute difference by minute,
    since the scale is not a concern there. 

\eat{
\vspace{-.15in}
{\small
\begin{eqnarray}
D(d)=\sqrt{{1 \over |\bar V(d)|}\sum_{v \in \bar V(d)}({v-v_0})^2}.
\label{eqn:deviation1}
\end{eqnarray}
}
\vspace{-.15in}
}
\end{itemize}

We have just defined dominant values, denoted by $v_0$. Regarding them, we also 
consider the following two measures.
\begin{itemize}\tightlist
  \item {\em Dominance factor:} The percentage of the sources
    that provide $v_0$ among all providers of $d$; that is,
    $F(d)={|\bar S(d,v_0)| \over |\bar S(d)|}$.
  \item {\em Precision of dominant values:} 
    The percentage of data items on which
    the dominant value is {\em true} (\ie, the same as the value in the gold standard).
\end{itemize}

\eat{
\begin{table}
\vspace{-.1in}
\centering
{\small
\caption{Default tolerance setting.
\label{tbl:tolerance_settings}}
\begin{tabular}{|c|c|}
\hline
Date type          & Default tolerance\\
\hline
Number  & Eq.(\ref{eqn:tolerance0}) with $\alpha$ = 0.01\\
Time      &   10-minute difference         \\
Text       &   Exactly the same ignoring case\\
\hline
\end{tabular}}
\vspace{-.1in}
\end{table}}
Before describing our results, we first clarify two issues 
regarding data processing. 
\begin{itemize}\tightlist 
  \item {\em Tolerance:} We wish to be fairly tolerant 
    to slightly different values. For time we are tolerant to 10-minute difference.
    For numerical values, we consider all values that are provided
    for each particular attribute $A$, denoted by $\bar V(A)$, and take the median;
    we are tolerant to a difference of 

\vspace{-.15in}
{\small
\begin{eqnarray}
\tau(A) = \alpha * \mbox{Median}(\bar V(A)),
\label{eqn:tolerance0}
\end{eqnarray}
}
\vspace{-.15in}

where $\alpha$ is a predefined {\em tolerance factor} and set to .01
by default.

  \item {\em Bucketing:} When we measure value distribution,
    we group values whose difference falls in our tolerance.
    Given numerical data item $d$ of attribute $A$,
    we start with the dominant value $v_0$, and
    have the following buckets: 
    $\dots,  (v_0-{3\tau(A) \over 2}, v_0-{\tau(A) \over 2}],
    (v_0-{\tau(A) \over 2}, v_0+{\tau(A) \over 2}],     
     (v_0+{\tau(A) \over 2}, v_0+{3\tau(A) \over 2}], \dots$.
\end{itemize}

\begin{figure*}[t]
\hspace{-.1in}
\begin{minipage}[b]{0.32\textwidth}
\includegraphics[width=2.3in]{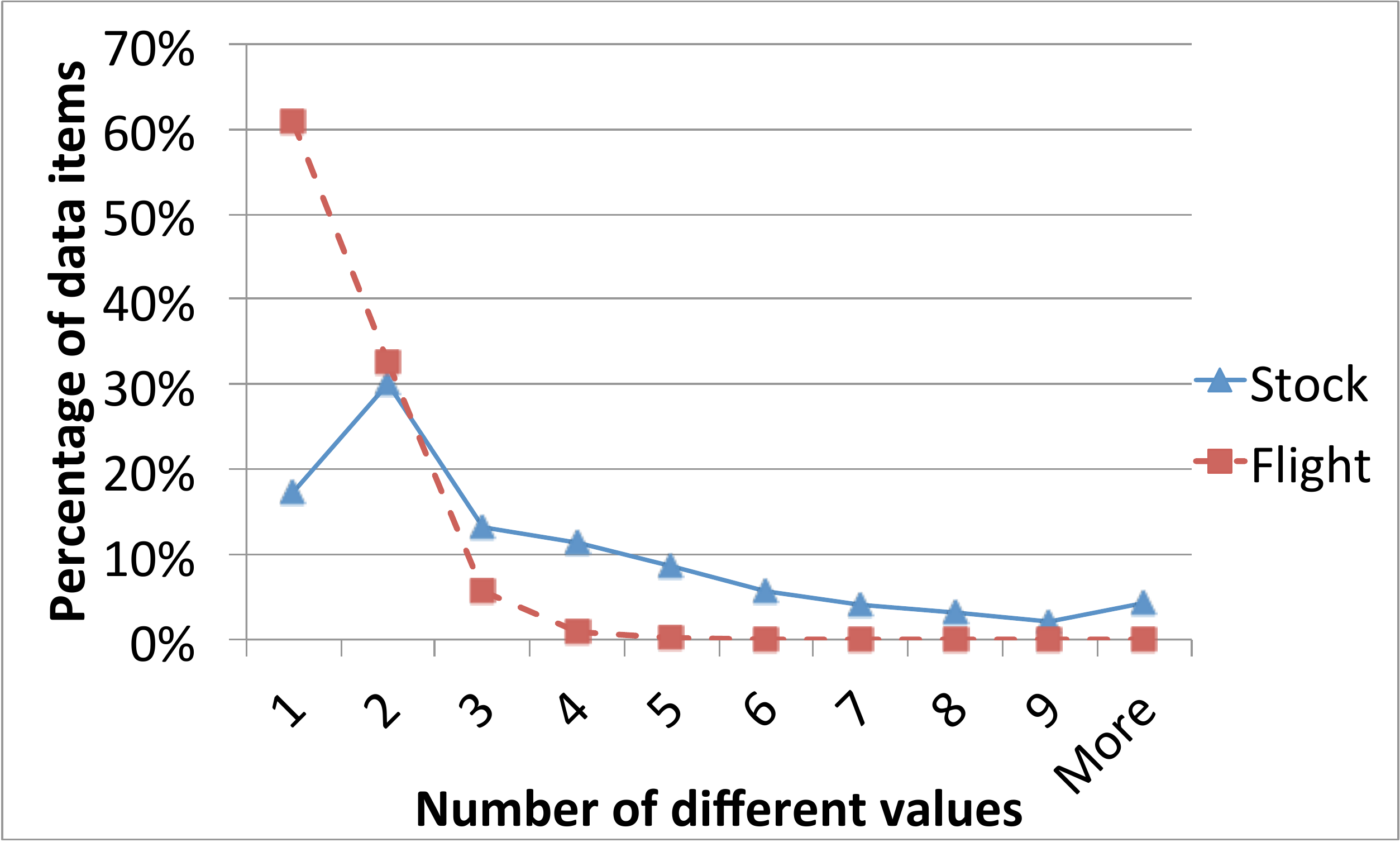}
\vspace{-.1in} 
\end{minipage}
\begin{minipage}[b]{0.32\textwidth}
\hspace{.1in}
\includegraphics[width=2.3in]{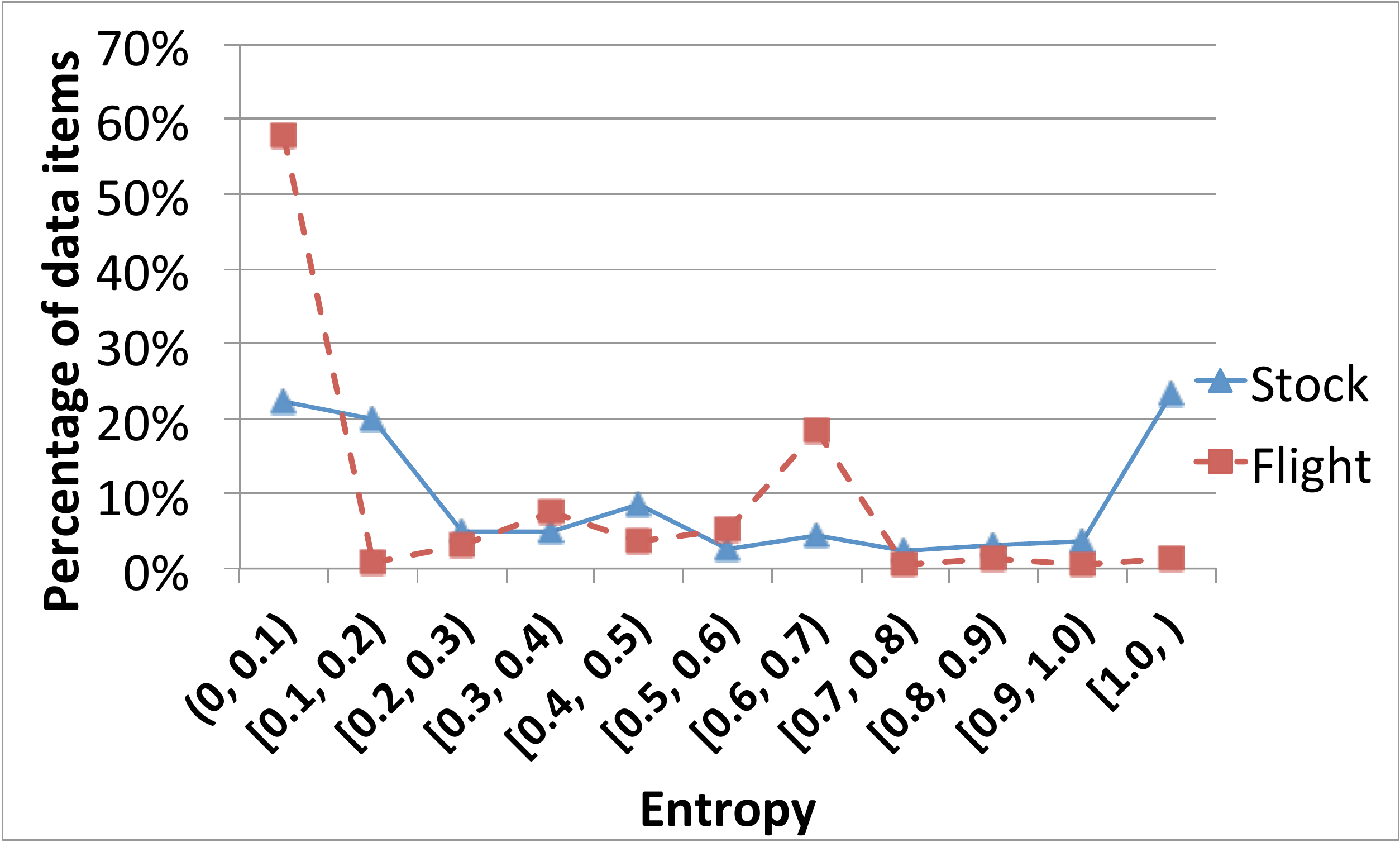}
\vspace{-.1in} 
\end{minipage}
\begin{minipage}[b]{0.32\textwidth}
\vspace{-1in}
\hspace{.2in}
\includegraphics[width=2.3in]{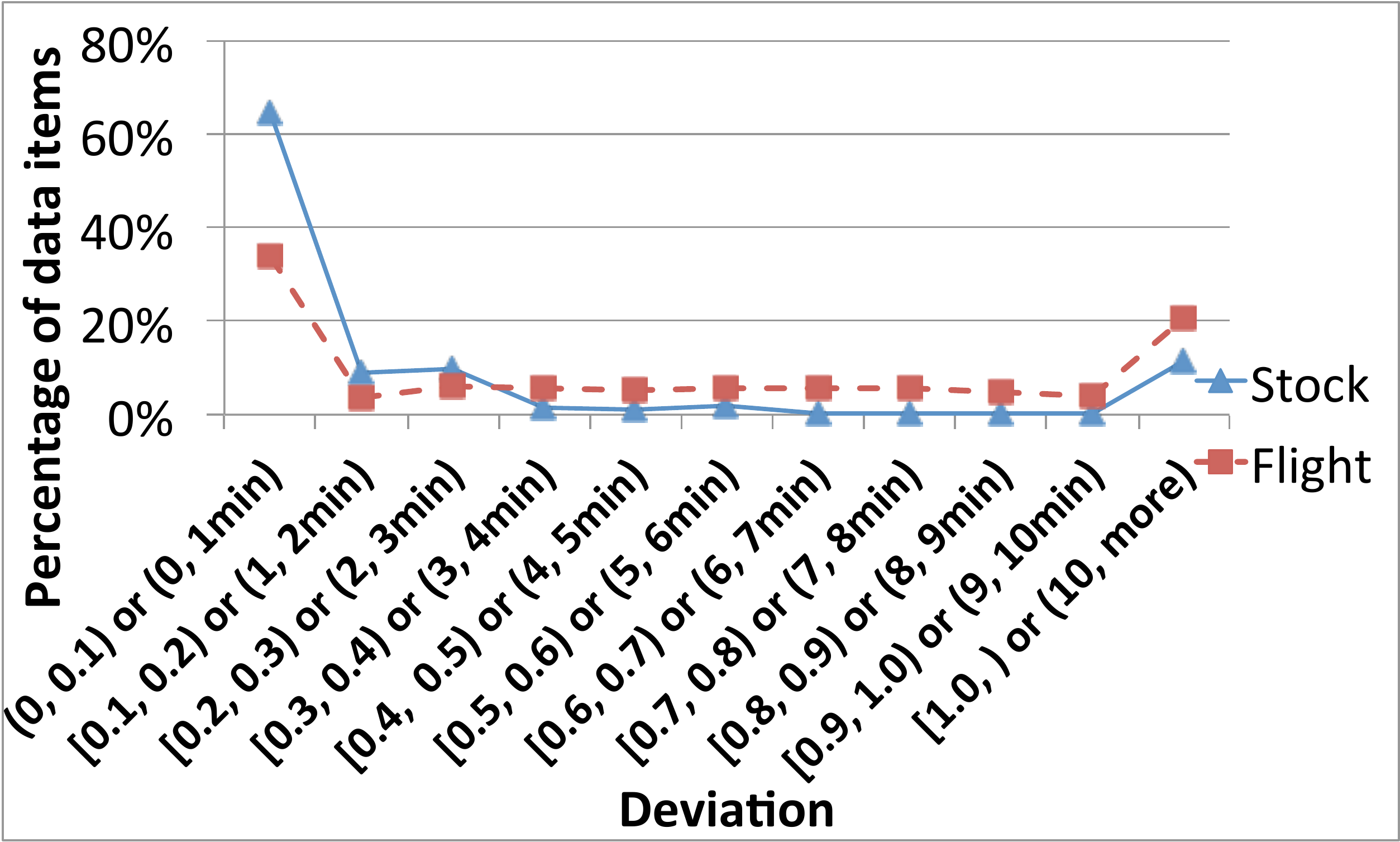}
\vspace{-.1in} 
\end{minipage}
\vspace{-.2in} 
\caption{Value inconsistency: distribution of number of values, entropy of
values, and deviation of numerical values.
\vspace{-.15in} 
\label{fig:value_variety}}
\end{figure*}
\begin{table}
\setlength{\tabcolsep}{4pt}
\vspace{-.1in}
\centering
{\small
\caption{Value inconsistency on attributes. The numbers in parentheses
are those when we exclude source {\em StockSmart}.}
\label{tbl:value_variety}
\begin{tabular}{|c|c|c|c|c|}
\hline
&Attribute w. & \multirow{2}{*}{Number} & Attribute w. & \multirow{2}{*}{Number} \\
& low incons. & & high incons. & \\
\hline
\multirow{5}{*} {\em Stock}& Previous close & 1.14 (1.14)& Volume & 7.42 (6.55)\\
&Today's high & 1.98 (1.18)& P/E & 6.89 (6.89)\\
&Today's low & 1.98 (1.18)&  Market cap  & 6.39 (6.39)\\
&Last price & 2.21 (1.33)& EPS  & 5.43 (5.43)\\
&Open price & 2.29 (1.29)& Yield & 4.85 (4.12)\\
\hline
\multirow{3}{*}{\em Flight} & Scheduled depart & 1.1  & Actual depart &  1.98\\
&Arrival gate& 1.18  &  Scheduled arrival &  1.65\\
&Depart gate& 1.19  &  Actual arrival &  1.6\\
\hline
\hline
&Low-var attr & Entropy & High-var attr & Entropy \\
\hline
\multirow{5}{*} {\em Stock}& Previous close & 0.04 (0.04)&  P/E & 1.49 (1.49)\\
&Today's high & 0.13 (0.05)& Market cap & 1.39 (1.39)\\
&Today's low & 0.13 (0.05) &   EPS  & 1.17 (1.17)\\
&Last price & 0.15 (0.07)& Volume & 1.02 (0.94)\\
&Open price & 0.19 (0.09)& Yield & 0.90 (0.90)\\
\hline
\multirow{3}{*}{\em Flight} & Scheduled depart & 0.05  & Actual depart &  0.60\\
&Depart gate& 0.10  &  Actual arrival &  0.31\\
&Arrival gate& 0.11  &  Scheduled arrival &  0.26\\
\hline
\hline
&Low-var attr & Deviation & High-var attr & Deviation \\
\hline
\multirow{5}{*} {\em Stock}& Last price &0.03 (0.02)& Volume & 2.96 (2.96)\\
&Yield & 0.18 (0.18)&52wk low price & 1.88 (1.88)\\
&Change \% & 0.19 (0.19)&   Dividend & 1.22 (1.22)\\
&Today's high & 0.33 (0.32)&  EPS & 0.81 (0.81)\\
&Today's low & 0.35 (0.33)&  P/E & 0.73 (0.73)\\
\hline
\multirow{2}{*}{\em Flight} & Schedule depart & 9.35 min & Actual depart&  15.14 min\\
&Schedule arrival& 12.76 min &  Actual arrival &  14.96 min\\
\hline
\end{tabular}}
\vspace{-.1in}
\end{table}

\smallskip
\noindent
{\bf Inconsistency of values:} Figure~\ref{fig:value_variety}
shows the distributions of inconsistency by different measures
for different domains and 
Table~\ref{tbl:value_variety} lists the attributes with the highest
or lowest inconsistency. 

\smallskip
\noindent
{\em Stock:} For the {\em Stock} domain, even with bucketing, 
the number of different values for a data item
ranges from 1 to 13, where the average is 3.7.  
There are only 17\% of the data items that have a single value,
the largest percentage of items (30\%) have two values, 
and 39\% have more than three values. 
However, we observe one source ({\em StockSmart})
that stopped refreshing data after June 1st, 2011;
if we exclude its data, 37\% data items have a single value,
16\% have two, and 36\% have more than three. 
The entropy shows that even though there are often multiple values,
very often one of them is dominant among others.
In fact, while we observe inconsistency on 83\% items, 
there are 42\% items whose entropy is less than
.2 and 76\% items whose entropy is less than 1
(recall that the maximum entropy for two values, happening
under uniform distribution, is 1).
After we exclude {\em StockSmart}, entropy on some attributes 
is even lower. 
Finally, we observe that for 64\% of the numerical data items
the deviation is within .1;
however, for 14\% of the items the deviation is above .5,
indicating a big discrepancy.
\eat{We did not observe that
the data from {\em StockSmart} affect the entropy on
statistical attributes and deviation much.}

The lists of highest- and lowest-inconsistency attributes
are consistent w.r.t. number-of-values and entropy,
with slight changes on the ordering. The lists w.r.t. deviation
are less consistent with the other lists. For some attributes
such as {\sf Dividend} and {\sf 52-week low price}, although there are 
not that many different values, the provided values can
differ a lot in the magnitude. Indeed, different sources can
apply different semantics for these two attributes:
{\sf Dividend} can be computed for different periods--year,
half-year, quarter, etc; {\sf 52-week low price} 
may or may not include the price of the current day. 
For {\sf Volume}, the high deviation is caused by 10 symbols
that have terminated--some sources map these symbols to other
symbols; for example, after termination of ``SYBASE'', symbol
``SY'' is mapped to ``SALVEPAR'' by a few sources.
When we remove these 10 symbols, the deviation drops to only .28.
Interestingly, {\sf Yield} has high
entropy but low deviation, because its values are typically 
quite small and the difference is also very small.
We observe that
real-time values often have a lower inconsistency than statistical values, 
because there is often more semantics ambiguity for statistical values.

\smallskip
\noindent
{\em Flight:} Value inconsistency is much lower for the {\em Flight} domain.
The number of different values ranges from 1 to 5 and the average is 1.45.
For 61\% of the data items there is a single value after bucketing
and for 93\% of the data items there are at most two values. 
There are 96\% of the items whose entropy is less than 1.0. 
However, when different times are provided for departure or arrival,
they can differ a lot: 46\% of the data items have a deviation above
5 minutes, while 20\% have a deviation above 10 minutes. 

Among different attributes, the scheduled departure time and 
gate information have the lowest
inconsistency, and as expected, the actual departure/arrival time
have the highest inconsistency. 
The average deviations for actual departure and arrival time are
as large as 15 minutes. 

\eat{
\begin{figure*}[t]
\begin{minipage}[b]{0.7\textwidth}
\includegraphics[width=3.3in]{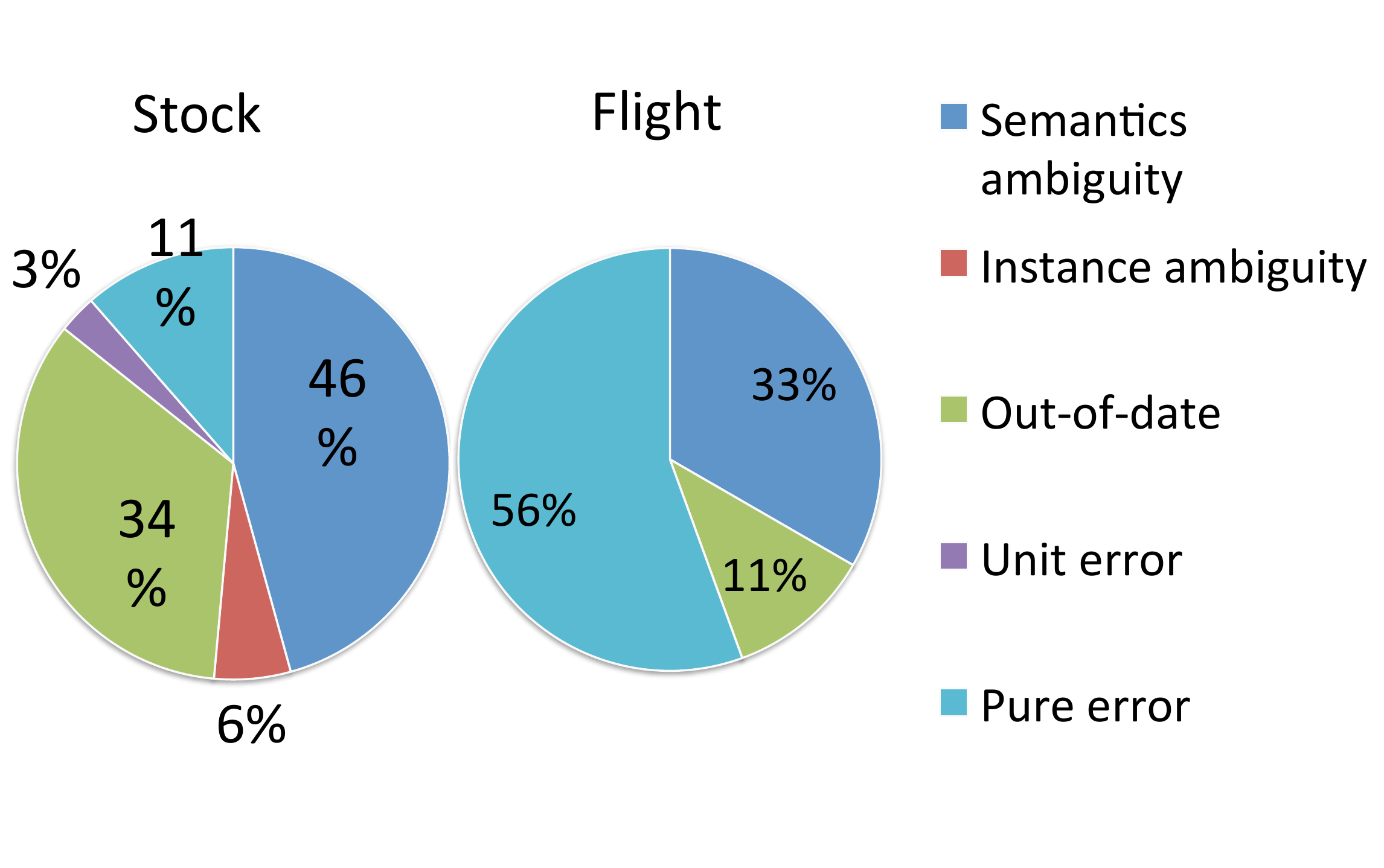}
\vspace{.0in}
\caption{Reasons for value inconsistency.\label{fig:varReason}}
\end{minipage}
\begin{minipage}[b]{0.28\textwidth}
\hspace{-1.6in}
\includegraphics[width=4in]{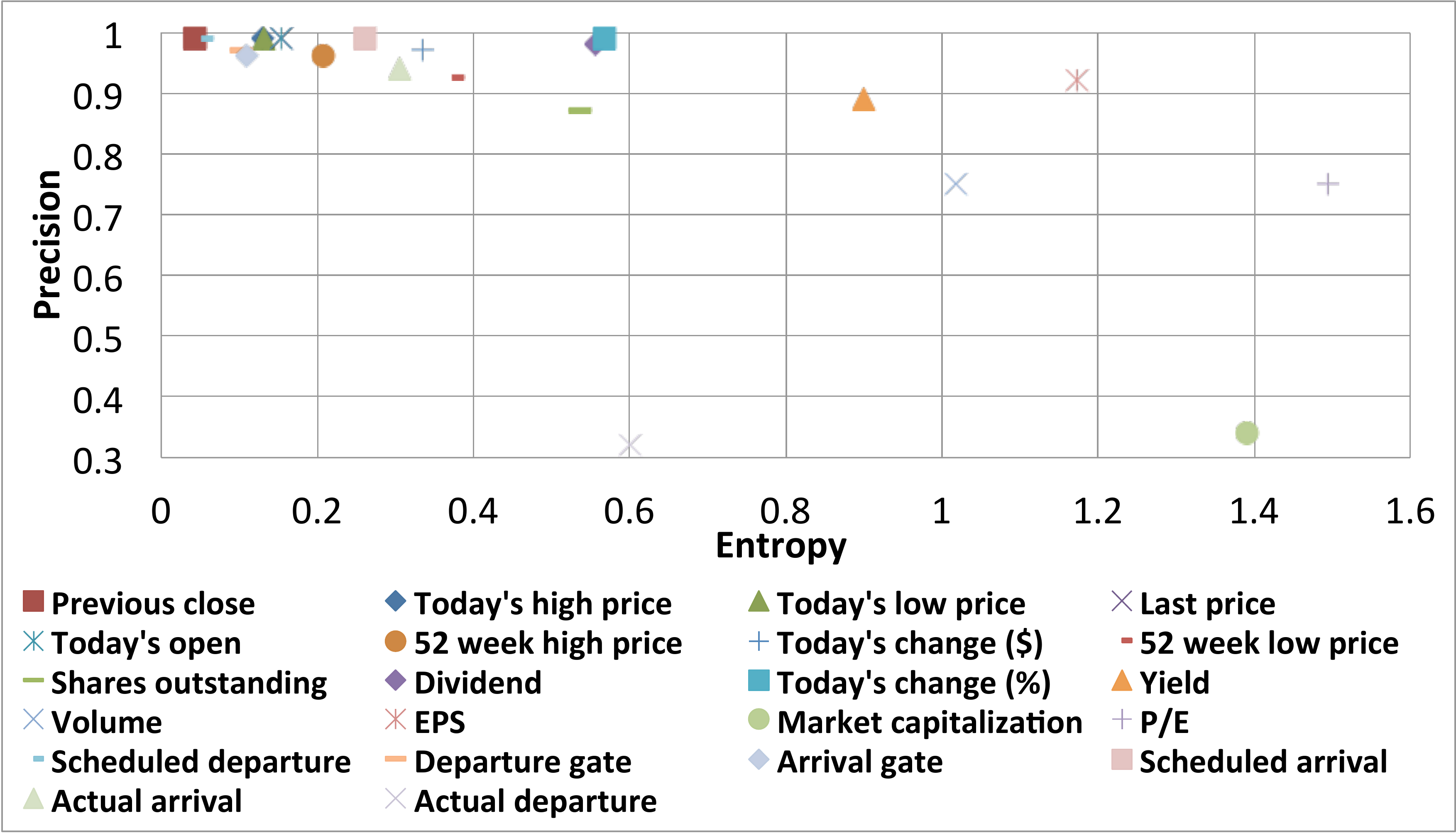}
\vspace{-.3in} 
\caption{Precision vs. entropy.\label{fig:correlate}}
\end{minipage}
\vspace{-.1in} 
\end{figure*}}
\smallskip
\noindent
{\bf Reasons for inconsistency:} To understand inconsistency of values,
for each domain we randomly chose 20 data items and in 
addition considered the 5 data items with the largest 
number-of-values, and
manually checked each of them to find the possible
reasons. Figure~\ref{fig:varReason} shows
the various reasons for different domains.

For the {\em Stock} domain, we observe five reasons.
(1) In many cases (46\%) the inconsistency is due to {\em semantics ambiguity}. 
We consider semantics ambiguity as the reason 
if ambiguity is possible for the particular 
attribute and we observe inconsistency between values provided by the
source and the dominant values on a large fraction
of items of that attribute; we have given examples of 
ambiguity for {\sf Dividend} and {\sf 52-week low price} earlier. 
(2) The reason can also be {\em instance ambiguity} (6\%),
where a source interprets one stock symbol differently from
the majority of sources; this happens mainly for
stock symbols that terminated at some point. Recall that
instance ambiguity results in the high deviation on {\sf Volume}.
(3) Another major reason is {\em out-of-date data} (34\%):
at the point when we collected data, the data were not up-to-date;
for two thirds of the cases the data were updated hours ago,
and for one third of the cases the data had not been refreshed for days.
(4) There is one error on data unit: the majority
reported 76M while one source reported 76B. 
(5) Finally, there are four cases (11\%) where we could not 
determine the reason and it seems to be {\em purely erroneous data}.

\begin{figure}
\centering
\includegraphics[width=3.4in]{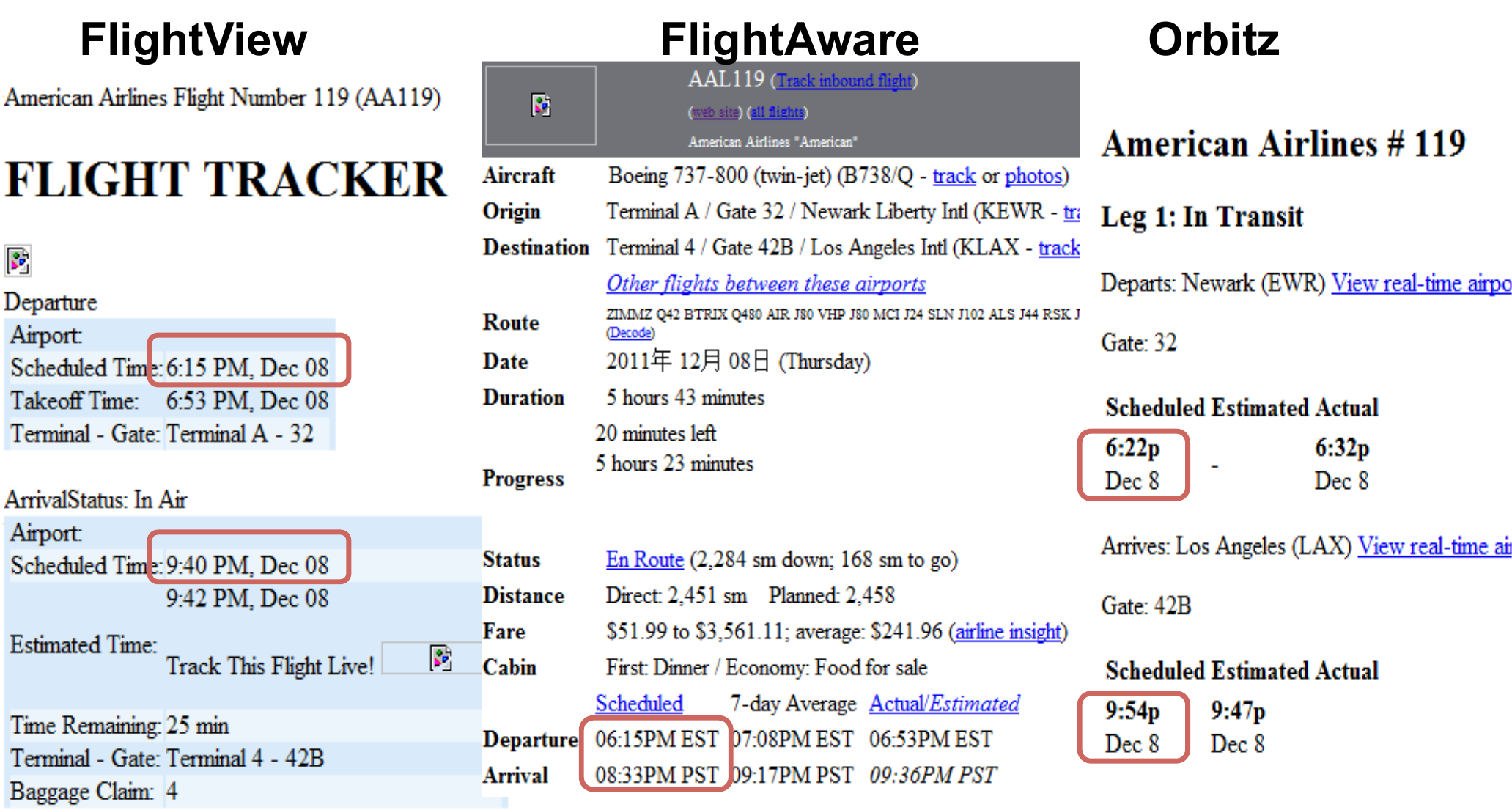}
\vspace{-.2in}
\small{
\caption{Screenshots of three flight sources.} 
\label{fig:flightScreenshot}}
\vspace{-.2in}
\end{figure}
For the {\em Flight} domain, we observe only three reasons.
(1) {\em Semantics ambiguity} causes 33\% of inconsistency:
some source may report takeoff
time as departure time and landing time as arrival time,
while most sources report the time of leaving the gate or arriving at the gate.
(2) {\em Out-of-date data} causes 11\% of the inconsistency; 
for example, even when a flight is already canceled,
a source might still report its actual departure and arrival time
(the latter is marked as ``estimated'').
(3) {\em Pure errors} seem to cause most of the inconsistency (56\%).
For example, Figure~\ref{fig:flightScreenshot}
shows three sources providing different scheduled departure
time and arrival time for Flight AA119 on 12/8/2011;
according to the airline website, the real scheduled time is 6:15pm
for departure and 9:40pm for arrival.
For scheduled departure time, {\em FlightView} and {\em FlightAware}
provide the correct time while {\em Orbitz} provides a wrong one.
For scheduled arrival time, all three sources provide different
times; {\em FlightView} again provides the correct one,
while the time provided by {\em FlightAware} is unreasonable
(it typically takes around 6 hours to fly from the east coast to
the west coast in the US). Indeed, we found that {\em FlightAware}
often gives wrong scheduled arrival time;
if we remove it, the average number of
values for {\em Scheduled arrival} drops from 1.65 to 1.31.

\begin{figure*}[t]
\begin{minipage}[b]{0.33\textwidth}
\includegraphics[width=2.2in]{figs/reasons.pdf}
\vspace{-.25in} 
\caption{Reasons for value inconsistency.\label{fig:varReason}}
\end{minipage}
\begin{minipage}[b]{0.66\textwidth}
\includegraphics[width=2.3in]{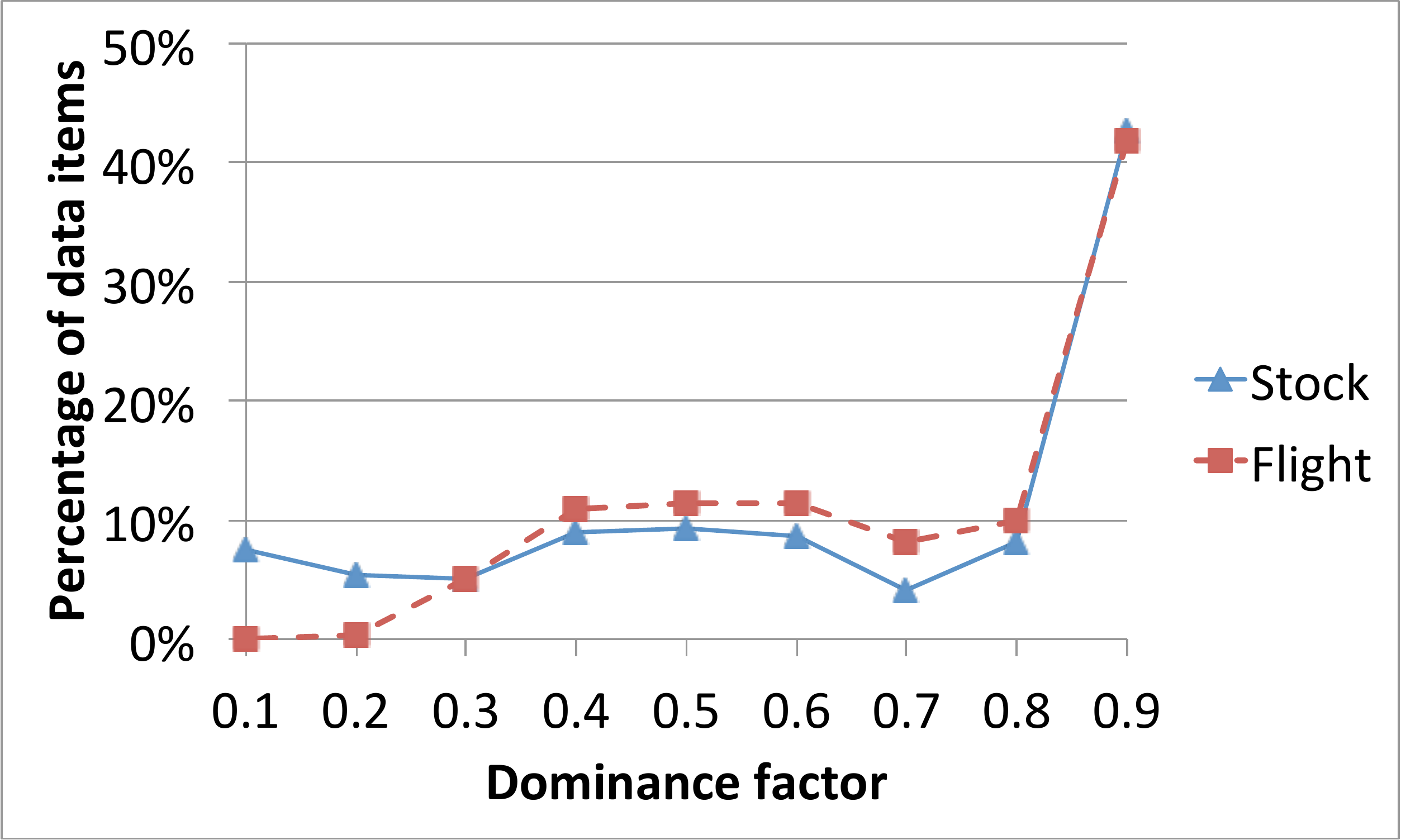}
\ \ 
\includegraphics[width=2.3in]{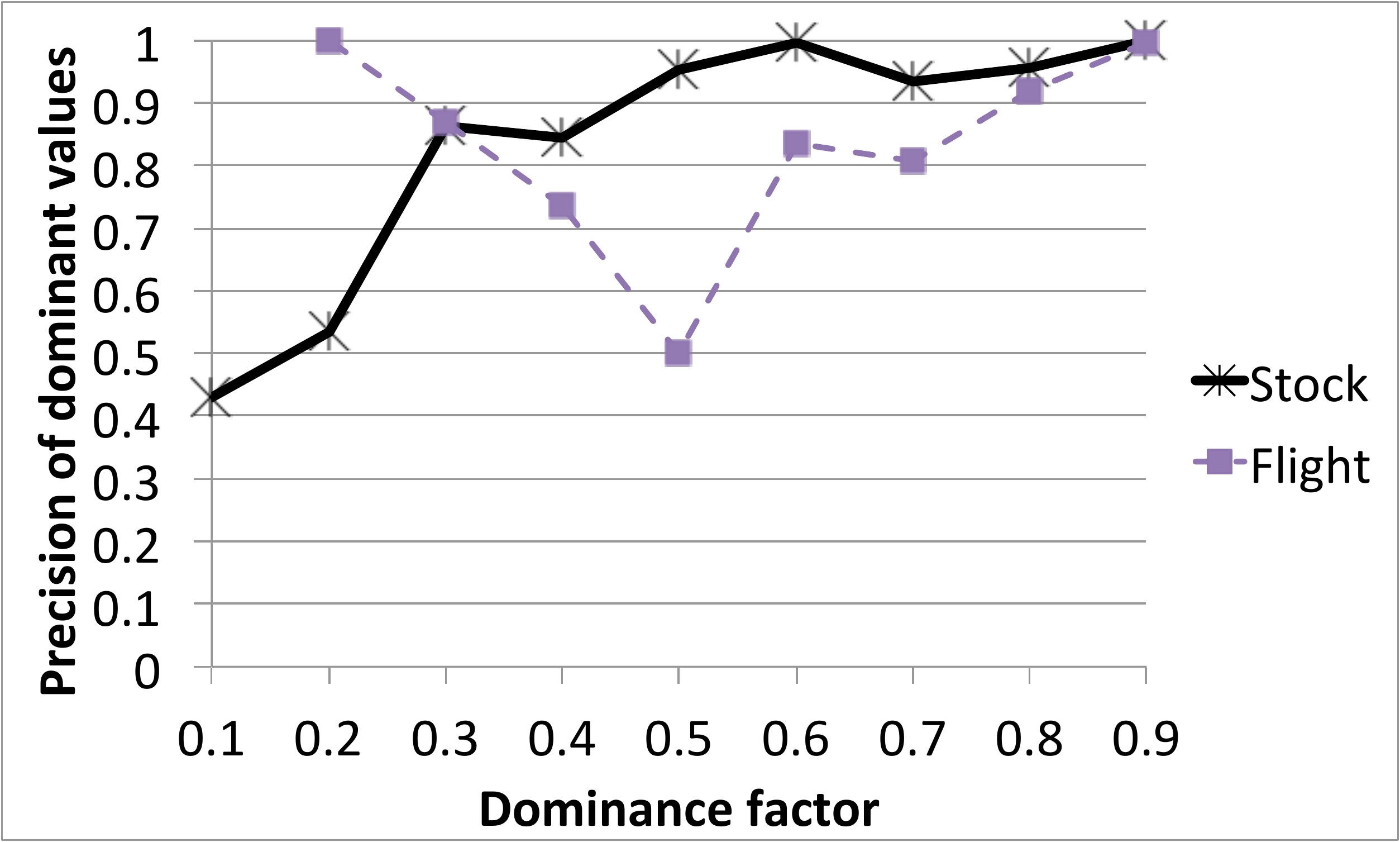}
\vspace{-.25in} 
\caption{Dominant values.\label{fig:dominance}}
\end{minipage}
\vspace{-.2in} 
\end{figure*}

\eat{
\begin{figure*}[t]
\begin{minipage}[b]{0.33\textwidth}
\includegraphics[width=2.4in]{figs/dominanceItems.pdf}
\vspace{-.1in} 
\end{minipage}
\begin{minipage}[b]{0.33\textwidth}
\hspace{0.1in}
\includegraphics[width=2.4in]{figs/dominantValuePrec.pdf}
\vspace{-.1in} 
\end{minipage}
\begin{minipage}[b]{0.33\textwidth}
\hspace{0.2in}
\includegraphics[width=2.4in]{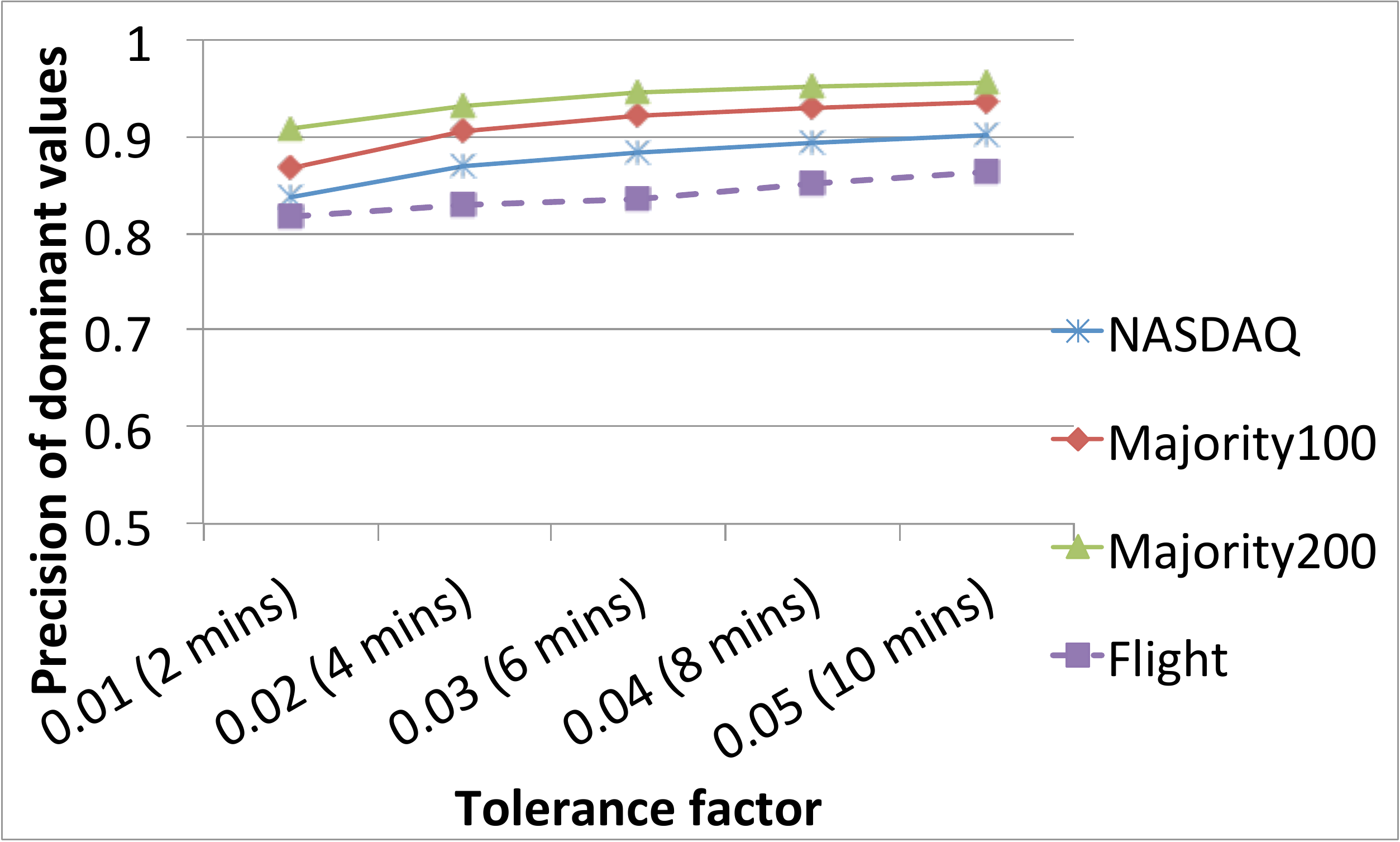}
\vspace{-.1in} 
\end{minipage}
\vspace{-.3in}
\caption{Dominant values: histogram on dominance factors, precision
of dominant values, and impact of tolerance factors. 
\label{fig:dominance}}
\vspace{-.1in}
\end{figure*}}

\smallskip
\noindent
{\bf Dominant values:} We now focus on the dominant values,
those with the largest number of providers for a given data item. 
Similarly, we can define
the {\em second dominant value}, etc. Figure~\ref{fig:dominance}
plots the distribution of the dominance factors
and the precision of the dominant values
with respect to different dominance factors.

For the {\em Stock} domain, we observe that on 42\% of the data items
the dominant values are supported by over 90\% of the sources,
and on 73\% of the data items the dominant values are supported by
over half of the sources. For these 73\% data items, 98\% of 
the dominant values are consistent with the gold standard.
However, when the dominance factor drops, the precision is also much lower.
For 9\% of the data items with dominance factor of .4, 
the consistency already drops to 84\%.
For 7\% of the data items
where the dominance factor is .1, the precision 
for the dominant value,
the second dominant value, and the third dominant value
is .43, .33, and .12 respectively (meaning that for 12\% of 
the data items none of the top-3 values is true).
\eat{In general, the precision w.r.t. {\em Majority200} is higher than
that w.r.t. {\em Majority100}, meaning a higher precision 
on the 100 symbols outside the NASDAQ index.
Also, the precision w.r.t. {\em Majority100} is higher than
that w.r.t. {\em NASDAQ}; indeed, we found that {\em NASDAQ}
contains 174 values that are not provided by
any other source on the same items.}
\eat{Finally, as we expected, when we increase the
tolerance factor $\alpha$ from .01 (reasonably strict) to 
.05 (quite tolerant), the precision increases 
from .91 to .96 w.r.t. {\em Majority200}.}

For the {\em Flight} domain, more data items have a higher dominance
factor--42\% data items have a dominance factor of over .9, and 82\%
have a dominance factor of over .5. However, for these 82\% items 
the dominant values have a lower precision: only 88\% are consistent
with the gold standard. Actually for the 11\% data items whose
dominance factor falls in $[.5,.6)$, the precision is only 50\% 
for the dominant value. As we show later, this is because 
some wrong values are copied between sources and become dominant.
\eat{As we increase the tolerance for time from 2 to 10 minutes,
the consistency for the dominant value increases smoothly
from .82 to .86. }

\eat{
Finally, Figure~\ref{fig:correlate} plots the entropy of the
attributes in various domains versus the precision of the
dominant values (for {\em Stock} we compared against 
{\em Majority200}). We observe
a tight correlation between these two measures: typically
the higher the entropy, the lower the precision of the dominant value.
This is as expected since a higher entropy often 
implies a lower dominance factor. 
We do not observe strong correlations between other 
consistency measures and the precision of dominant values.
}

\smallskip
\noindent
{\bf Summary and comparison:} Overall we observe a fairly high 
inconsistency of values on the same data item:
for {\em Stock} and {\em Flight} 
the average entropy is .58 and .24, and the average deviation
is 13.4 and 13.1 respectively.
The inconsistency can vary from attributes to attributes.
There are different reasons for the inconsistency, including ambiguity,
out-of-date data, and pure errors. 
For the {\em Stock} domain, half of the inconsistency
is because of ambiguity, one third is because of out-of-date
data, and the rest is because of erroneous data.
For the {\em Flight} domain, 56\% of the inconsistency is
because of erroneous data.

If we choose dominant values as the true value
(this is essentially the {\sc Vote} strategy, as
we explain in Section~\ref{sec:fusion}), we can obtain
a precision of 0.908 for {\em Stock} 
and 0.864 for {\em Flight}.
We observe that dominant values with a high
dominance factor are typically correct, but the precision
can quickly drop when this factor decreases.
Interestingly, the {\em Flight} domain has a lower inconsistency
but meanwhile a lower precision for dominant values, mainly
because of copying on wrong values, as we show later.

\subsection{Source accuracy}
Next, we examine the accuracy of the sources over time.
Given a source $S$, we consider the following two measures.
\begin{itemize}\tightlist
  \item {\em Source accuracy:} We compute accuracy of $S$
    as the percentage of its provided true values among all its
    data items appearing in the gold standard.
  \item {\em Accuracy deviation:} We compute the standard deviation
    of the accuracy of $S$ over a period of time. 
    We denote by $\bar T$ the time points in a period, 
    by $A(t)$ the accuracy of $S$
    at time $t \in \bar T$, and by $\hat A$ the mean accuracy over 
    $\bar T$. The variety is computed by 
    $\sqrt{{1 \over |\bar T|}\sum_{t \in \bar T}(A(t)-\hat A)^2}$.
\end{itemize}

\begin{figure*}[t]
\vspace{.1in}
\begin{minipage}[b]{0.33\textwidth}
\center
\includegraphics[width=2.2in]{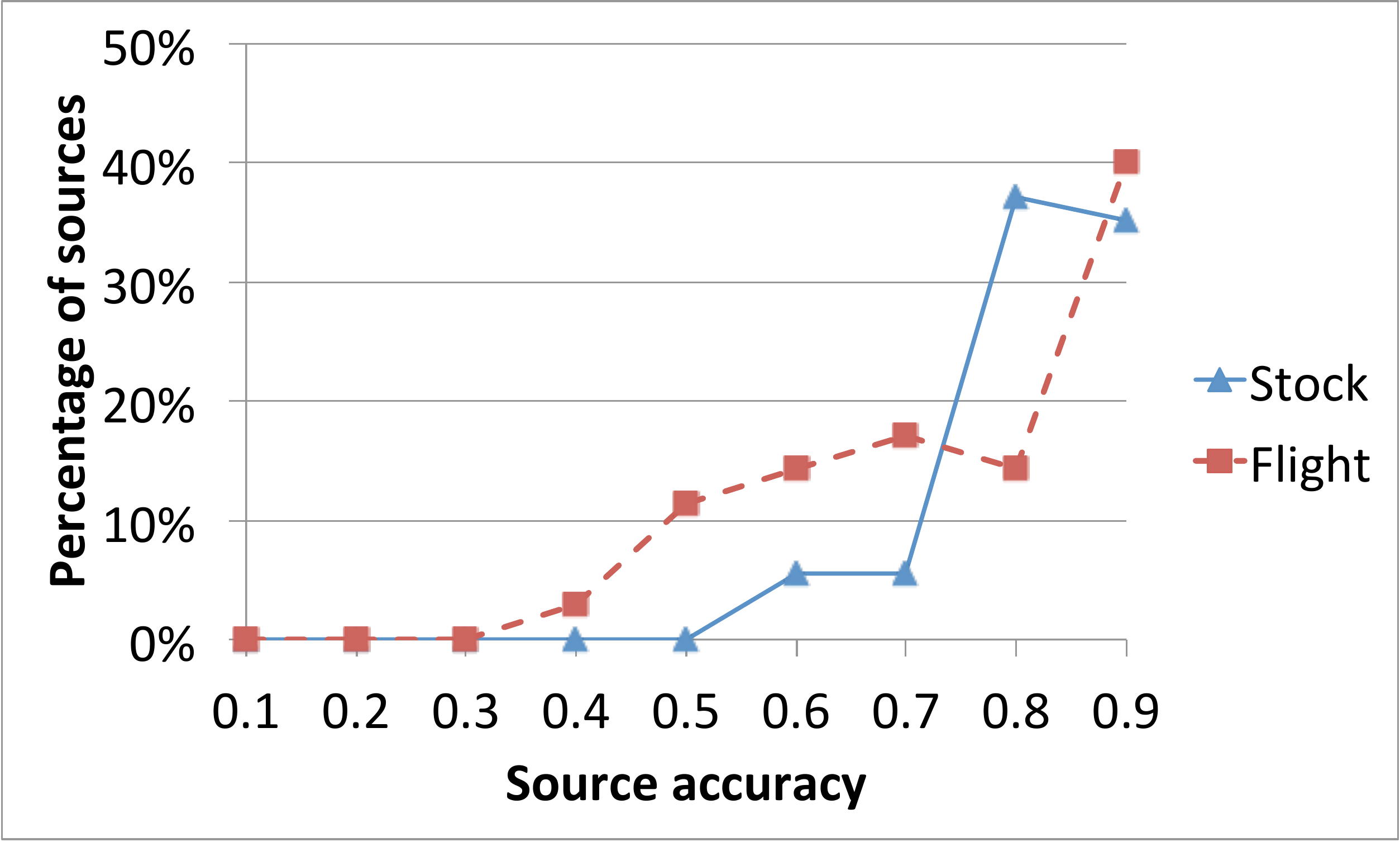}

(a) Distribution of source accuracy.

\end{minipage}
\begin{minipage}[b]{0.33\textwidth}
\center
\includegraphics[width=2.2in]{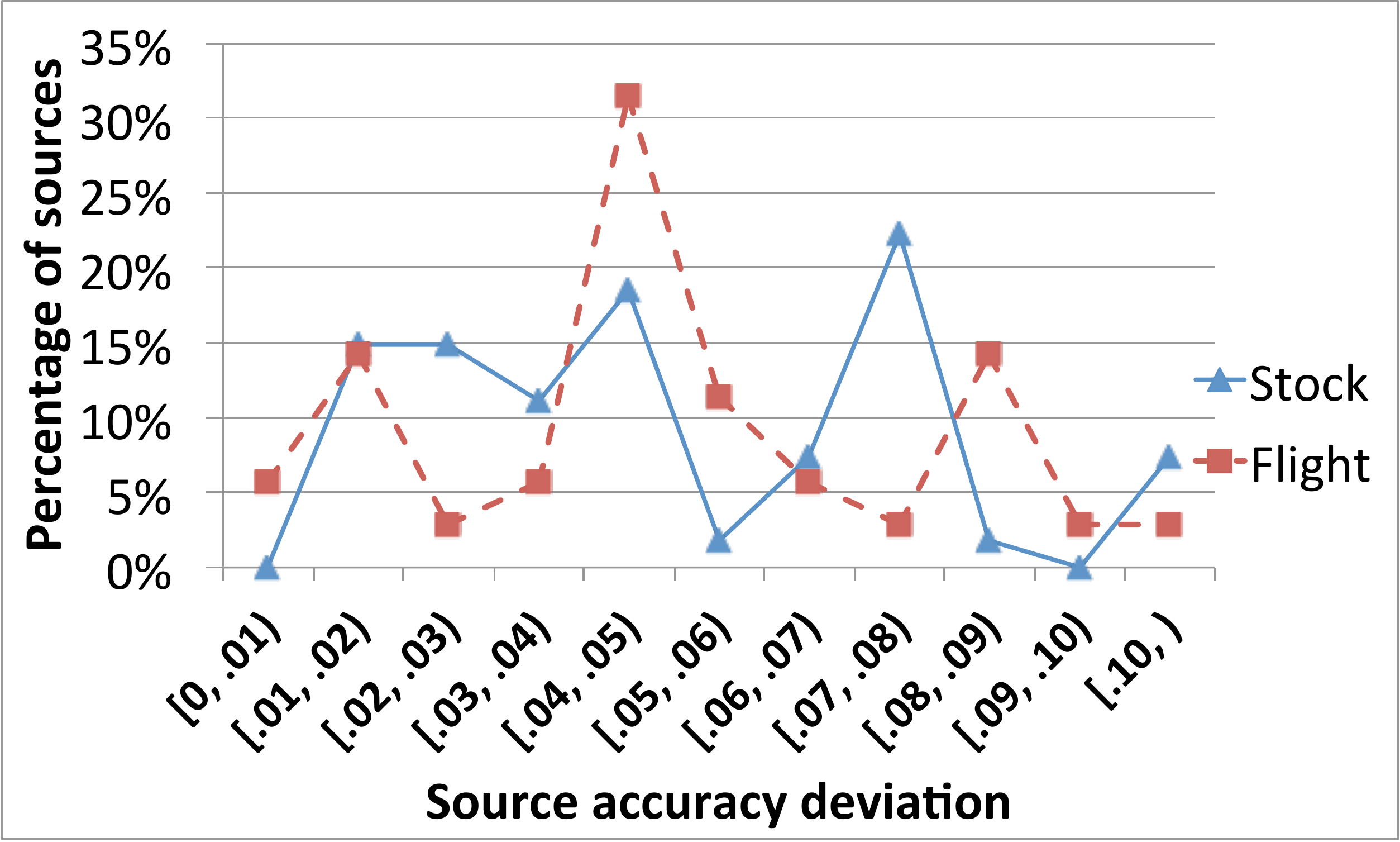}
(b) Accuracy deviation over time.

\end{minipage}
\begin{minipage}[b]{0.33\textwidth}
\center
\includegraphics[width=2.2in]{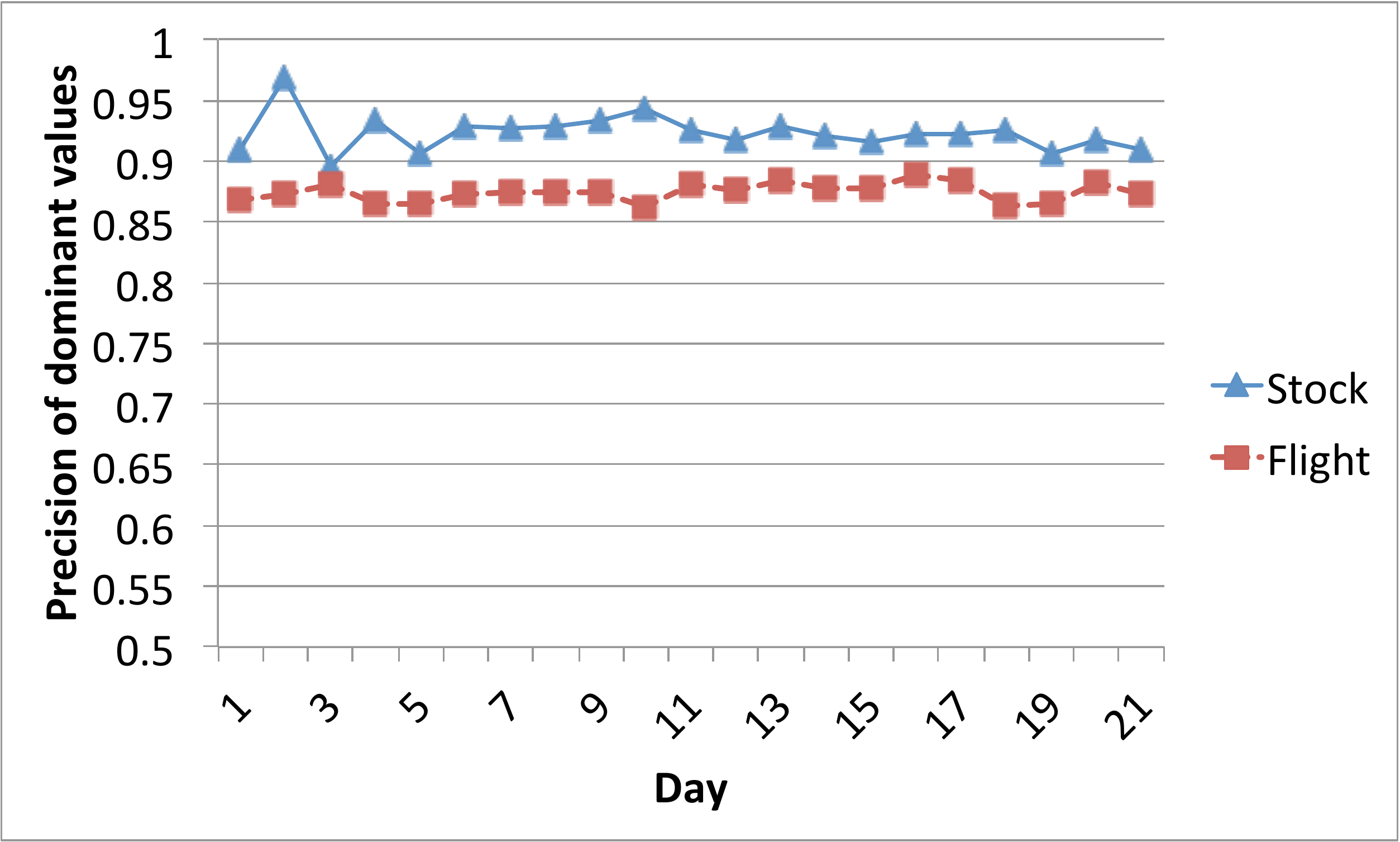}
(c) Dominant values over time.

\end{minipage}
\vspace{-.25in} 
\caption{Source accuracy and deviation over time.
\label{fig:accuracy}}
\vspace{-.15in} 
\end{figure*}

\eat{
\begin{figure*}[t]
\vspace{-.5in}
\begin{minipage}[b]{0.33\textwidth}
\includegraphics[width=2.4in]{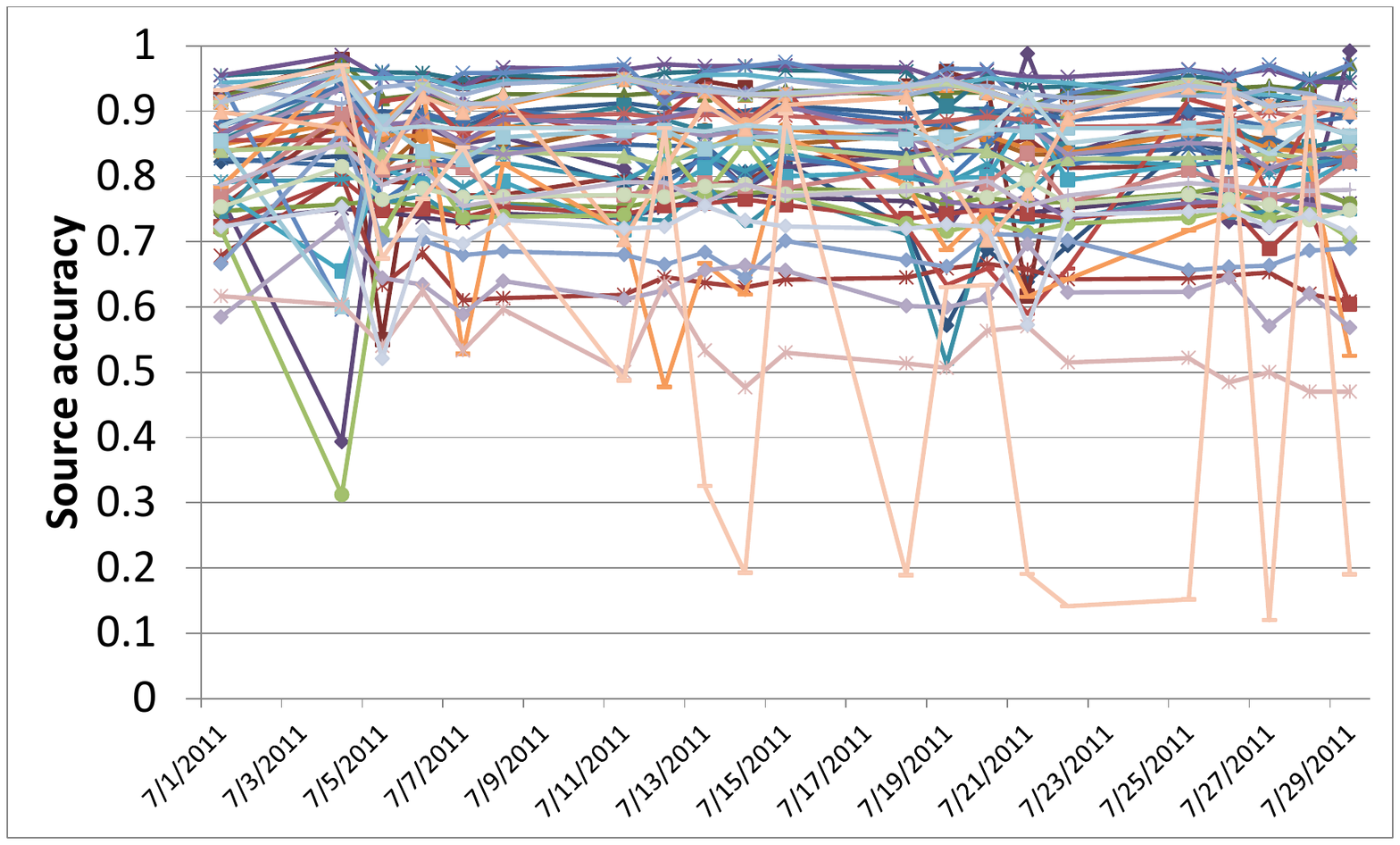}
\vspace{-.1in} 
\end{minipage}
\begin{minipage}[b]{0.33\textwidth}
\hspace{-.1in}
\includegraphics[width=2.4in]{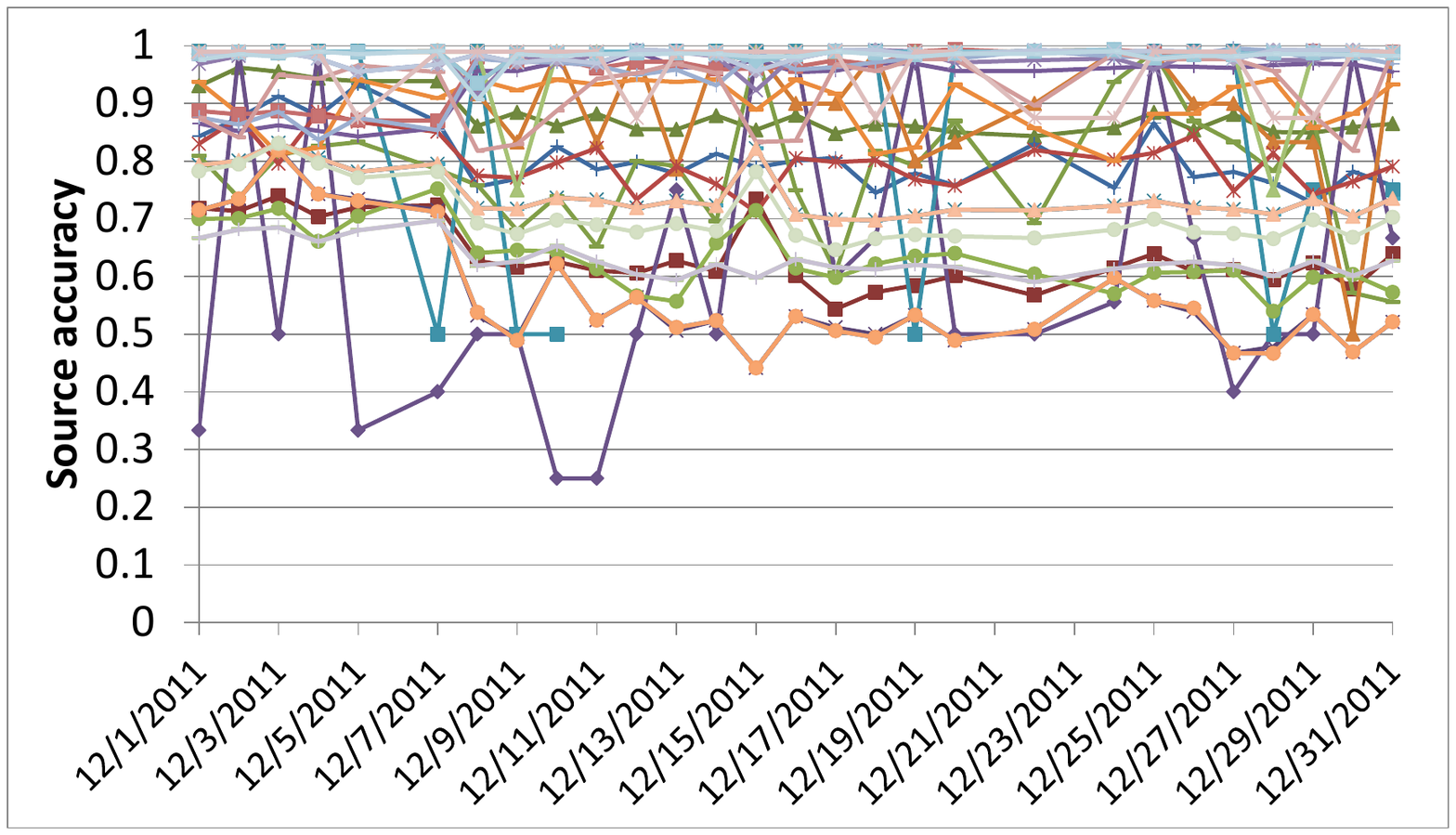}
\vspace{-.1in} 
\end{minipage}
\begin{minipage}[b]{0.33\textwidth}
\includegraphics[width=2.4in]{figs/flightSourceAccuOverTime.pdf}
\vspace{-.1in} 
\end{minipage}
\vspace{-1.2in} 
\caption{Source accuracy over one month on each domain
\label{fig:steadiness}}
\end{figure*}
}

\smallskip
\noindent
{\bf Source accuracy:} Figure~\ref{fig:accuracy}(a) shows the 
distribution of source accuracy in 
different domains. Table~\ref{tbl:authoratative} lists the
accuracy and item-level coverage of some authoritative sources. 

In the {\em Stock} domain, the accuracy varies from .54 to
.97 (except {\em StockSmart}, which has accuracy .06), 
with an average of .86. Only 35\% sources have
an accuracy above .9, and 3 sources (5\%) have an accuracy 
below .7, which is quite low. Among the five popular 
financial sources, four have an accuracy above .9, 
but {\em Bloomberg} has an accuracy of only .83 because
it may apply different semantics on some statistical attributes
such as {\sf EPS, P/E} and {\sf Yield}. All
authoritative sources have a coverage between .8 and .9.

In the {\em Flight} domain, we consider sources excluding 
the three official airline websites (their data are used
as gold standard). The accuracy varies from .43 to
.99, with an average of .80. There are 40\% of the sources with
an accuracy above .9, but 10 sources (29\%) have
an accuracy below .7. 
The average accuracy of airport sources is .94, but their
average coverage is only .03. 
Authoritative sources like {\em Orbitz} and {\em Travelocity} 
all have quite high accuracy (above .9), but {\em Travelocity} 
has low coverage (.71).

\begin{table}
\centering
{\small
\caption{Accuracy and coverage of authoritative sources.}
\label{tbl:authoratative}
\begin{tabular}{|c|c|c|c|}
\hline
 & Source & Accuracy & Coverage \\
\hline
\multirow{5}{*}{Stock} & {\em Google Finance} & .94 & .82\\
& {\em Yahoo! Finance} & .93 &.81\\
& {\em NASDAQ} & .92 & .84\\
& {\em MSN Money}  & .91 &.89\\
& {\em Bloomberg} & .83& .81\\
\hline
\multirow{3}{*}{Flight}& {\em Orbitz} & .98 & .87\\
& {\em Travelocity} & .95 &  .71\\
& {\em Airport average}  & .94 & .03\\
\hline
\end{tabular}}
\vspace{-.2in} 
\end{table}

\smallskip
\noindent
{\bf Accuracy deviation:} Figure~\ref{fig:accuracy}(b) shows
the accuracy deviation of the sources
in a one-month period, and Figure~\ref{fig:accuracy}(c)
shows the precision of the dominant values over time. 

In the {\em Stock} domain, we observe that for 4 sources
the accuracy varies tremendously (standard deviation over .1)
and the highest standard deviation is as high as .33. For 59\% of the sources 
the accuracy is quite steady (standard deviation below .05).
We did not observe any common peaks or dips on particular days.
The precision of the dominant values 
ranges from .9 to .97, and the average is .92.
The day-by-day precision is also fairly smooth,
with some exceptions on a few days.

In the {\em Flight} domain, we observe that for 1 source
the accuracy varies tremendously (deviation .11),
and for 60\% sources the accuracy is quite steady (deviation below .05).
The precision of the dominant values
ranges from .86 to .89, and the average is .87.

\smallskip
\noindent
{\bf Summary and comparison:} We observe that the accuracy of
the sources can vary a lot. On average the accuracy 
is not too high: .86 for {\em Stock} and .80 for {\em Flight}. 
Even authoritative sources may not have very high accuracy. 
We also observe that the accuracy is fairly steady in general.
On average the standard deviation is 0.06 for {\em Stock} and 
0.05 for {\em Flight}, and for about half of the sources the deviation
is below .05 over time.

\eat{
We crawled data from each data sources for a long period on daily basis. In this section, we study the data consistency of each data sources over time to get a better understanding of source quality. Besides, the consistency results would provide evidence for evaluating the robustness of each fusion methods.

\begin{figure}
\centering
\includegraphics[width=2.5in]{figs/source_accuracy_consistency.pdf}
\vspace{-.2in}
\caption{Source Accuracy Variance}\label{fig:source_accuracy_consistency}
\end{figure}
}

\eat{
\begin{table}
\centering
{\small
\caption{Precision of dominant values with and without 
potential copiers.\label{tbl:voteOnDepen}}
\begin{tabular}{|c|c|c|}
\hline
     & {\em Stock} & {\em Flight} \\
\hline
W. copiers & .908 & .864 \\
W/o. copiers & .923 & .927\\
\hline
\end{tabular}}
\vspace{-.2in} 
\end{table}
}

\eat{
\begin{figure}
\centering
\includegraphics[width=2.5in]{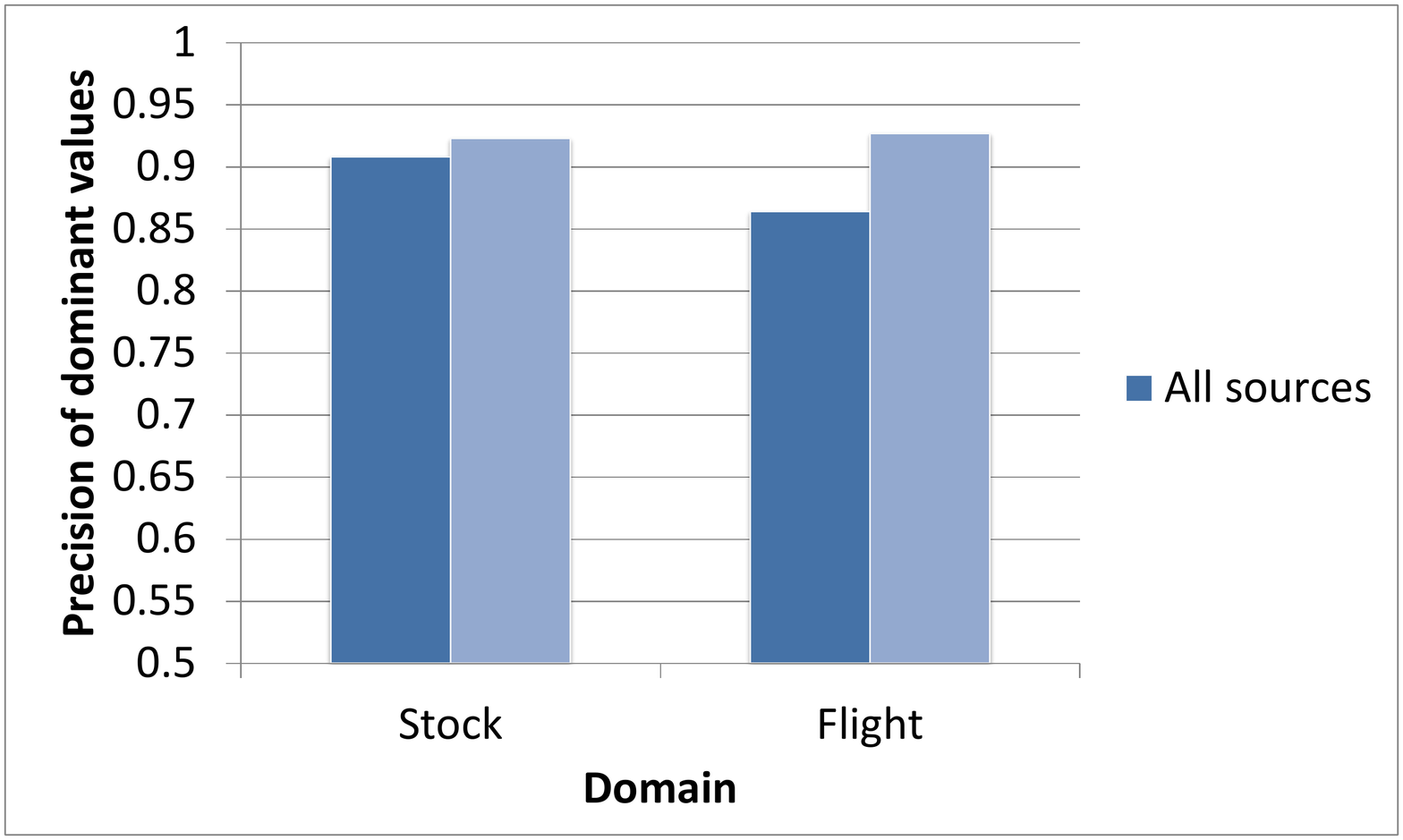}
\caption{Precision of dominant values with and without 
potential copiers.
\label{fig:voteOnDepen}}
\end{figure}}

\subsection{Potential copying}
\label{sec:copying}
Just as copying is common between webpage texts,
blogs, etc., we also observe copying
between deep-web sources; that is, one source obtains
some or all of its data from another source, while
possibly adding some new data independently. 
We next report the potential copying we found in our
data collections (Table~\ref{tbl:depenSources}) 
and study how that would affect precision
of the dominant values. 
For each group $\bar S$ of sources with copying, we compute the
following measures.

\begin{itemize}\tightlist
  \item {\em Schema commonality:} We measure schema commonality
    as the average Jaccard similarity between the
    sets of provided attributes on each pair of sources.
    If we denote by $\bar A(S)$ the set of global attributes
    that $S$ provides, we compute
    schema commonality of $\bar S$ as 
    $Avg_{S, S' \in \bar S, S \ne S'}{|\bar A(S) \cap \bar A(S')| \over
      |\bar A(S) \cup \bar A(S')|}$.
  \item {\em Object commonality:} Object commonality is also
    measured by average Jaccard similarity but between the sets
    of provided objects. 
  \item {\em Value commonality:} The average percentage of common values 
    over all shared data items between each source pair.
  \item {\em Average accuracy:} The average source accuracy.
\end{itemize}

\eat{
\begin{figure}
\vspace{.3in}
\includegraphics[width=3in]{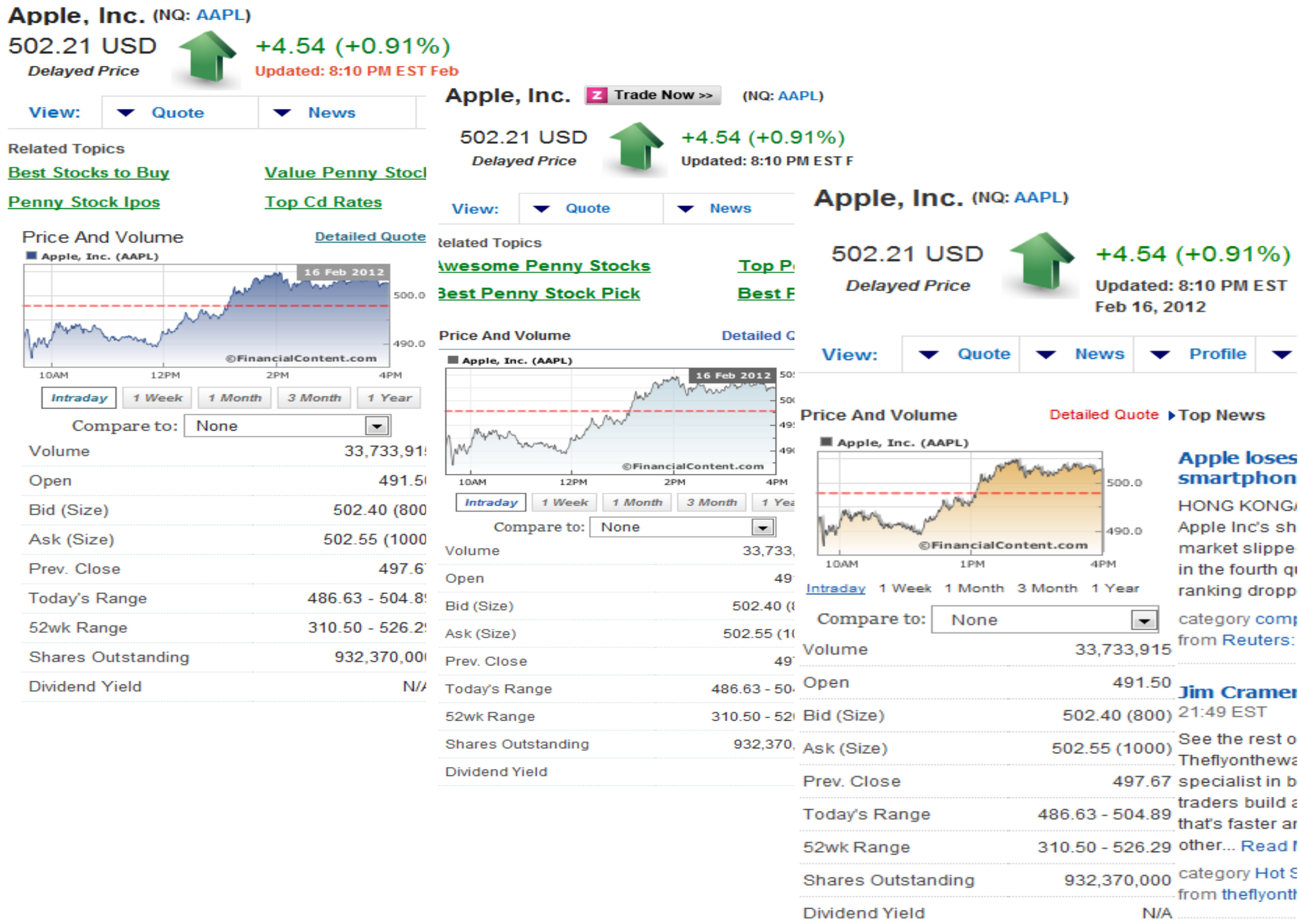}
\small{
\caption{Screenshots of three stock sources.} 
\label{fig:depenPage}}
\vspace{-.2in} 
\end{figure}
}
On the {\em Stock} domain, we found two groups of sources
with potential copying. The first group contains 11 sources,
with exactly the same webpage layout, schema, and highly similar
data. 
These sources all derive their data from {\em Financial Content}, 
a market data service company, and their data are quite accurate 
(.92 accuracy). The second group contains 2 sources,
also with exactly the same schema and data; the two websites are indeed 
claimed to be merged in 2009. However, their data have an accuracy
of only .75. For each group, we keep only
one randomly selected source and remove the rest of the sources;
this would increase the precision of dominant values from .908
to .923. 

On the {\em Flight} domain, we found five groups of sources
with potential copying. Among them, two directly claim 
partnership by including the logo of other sources; one re-directs
its queries; and two embed the query interface of other sources.
Sources in the largest two groups provide a little bit different
sets of attributes, but exactly the same flights, and the same
data for all overlapping data items. Sources in other groups 
provide almost the same schema and data. Accuracy of sources
in these groups vary from .53 to .93.
After we removed the copiers and kept only one
randomly selected source in each group, the precision of
dominant values is increased significantly, from .864 to .927.

\smallskip
\noindent
{\bf Summary and comparison:} We do observe copying between
deep-web sources in each domain. In some cases the copying
is claimed explicitly, and in other cases it is detected
by observing embedded interface or query redirection.
For the copying that we have observed, while the sources may
provide slightly different schemas, they provide almost the
same objects and the same values.
The accuracy of the original sources may not be high, 
ranging from .75 to .92 for {\em Stock}, and from
.53 to .93 for {\em Flight}. Because the {\em Flight}
domain contains more low-accuracy sources with copying,
removing the copied sources improves the precision of 
the dominant values more significantly than in the {\em Stock}
domain.

\begin{table}
\setlength{\tabcolsep}{4pt}
\centering
{\small
\caption{Potential copying between sources.
\label{tbl:depenSources}}
\begin{tabular}{|c|c|c|c|c|c|c|}
\hline
     & \multirow{2}{*}{Remarks} & \multirow{2}{*}{Size} & Schema & Object & Value & Avg\\
     &      &         & sim & sim & sim & accu\\
\hline
\multirow{2}{*}{Stock}& Depen claimed & 11 & 1 & .99 & .99 & .92\\
 & Depen claimed & 2 & 1 & 1 & .99 & .75    \\
\hline
\multirow{5}{*}{Flight} & Depen claimed & 5 & 0.80 & 1 & 1 &.71\\
& Query redirection  & 4 & 0.83  & 1 & 1 & .53\\
& Depen claimed & 3 & 1 & 1 & 1 & .92\\
& Embedded interface & 2 & 1 & 1 & 1 & .93\\
& Embedded interface & 2 & 1 & 1 & 1 & .61\\
\hline
\end{tabular}}
\vspace{-.2in} 
\end{table}

\section{Data Fusion}
\label{sec:fusion}
\eat{
\begin{figure*}[t]
\vspace{-.2in}
\hspace{-.2in}
\includegraphics[width=7in]{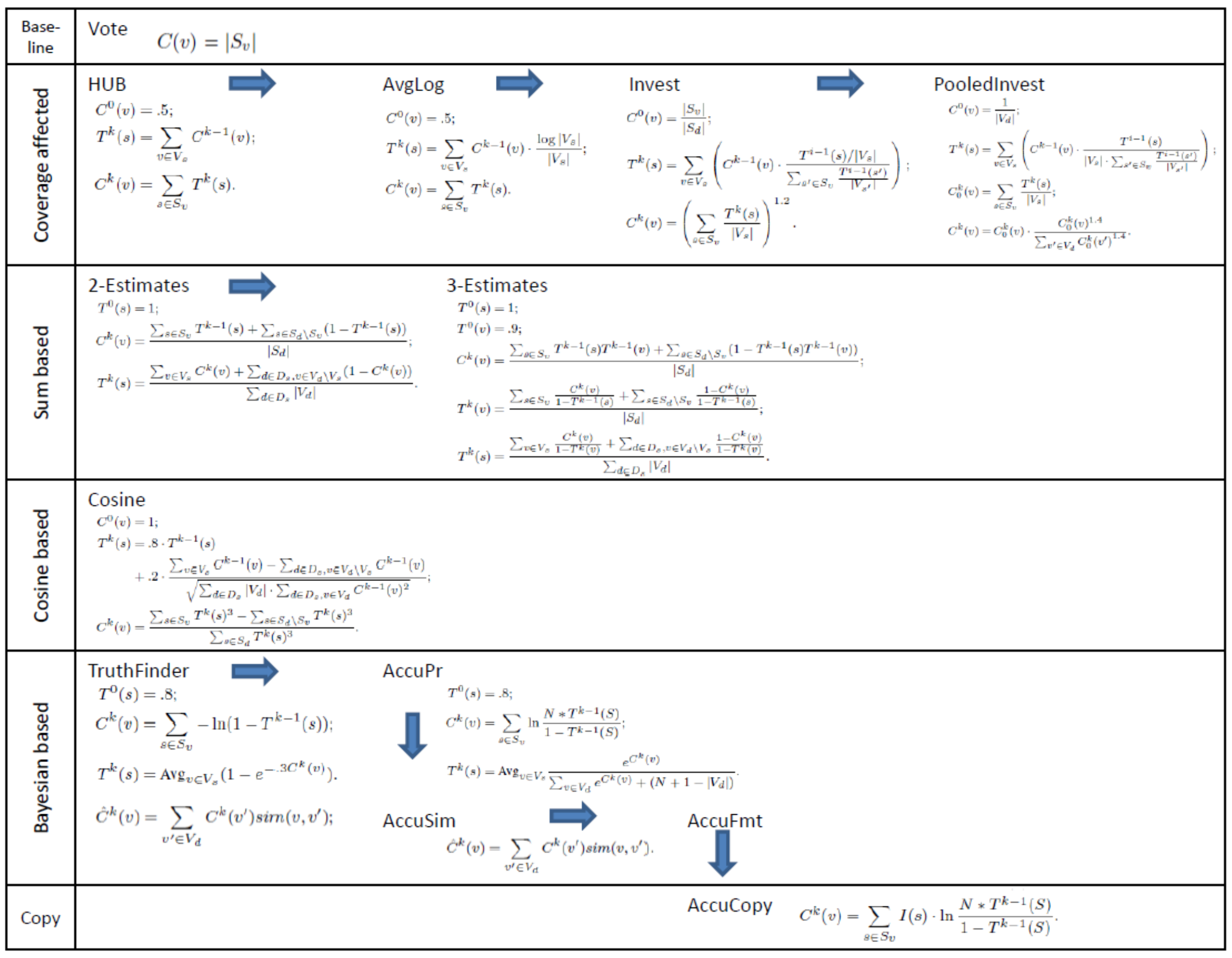}
\vspace{-.2in} 
\caption{\small Various fusion methods and their relationships.\label{fig:fusion}}
\vspace{-.1in} 
\end{figure*}
}

As we have shown in Section~\ref{sec:quality}, deep-web data from different
sources can vary significantly and there can be a lot of conflicts.
{\em Data fusion} aims at resolving conflicts and finding the true values.
A basic fusion strategy that considers the dominant value (\ie, the value with the 
largest number of providers) as the truth works well when the dominant
value is provided by a large percentage of sources (\ie, a high dominance factor),
but fails quite often otherwise. Recall that in the {\em Stock} domain,
the precision of dominant values is 90.8\%, meaning that on around
1500 data items we would conclude with wrong values. 
Recently many advanced fusion techniques have been proposed to improve
the precision of truth discovery~\cite{BCM+10, BN08, DBS09a, 
DBS09b, DN09, GAMS10, PR10, PR11, WM07, WM11, YHY08, YT11, ZRHG12}. 

In this section we answer the following three questions.
\begin{enumerate}\tightlist
  \item {\em Are the advanced fusion techniques effective?}
    In other words, do they perform (significantly) better 
    than simply taking the dominant values or taking all data
    provided by the best source (assuming we know which 
    source it is).
  \item {\em Which fusion method is the best?} In other words,
    is there a method that works better than others on all
    or most data sets?
  \item {\em Which intuitions for fusion are effective?} 
    In other words, does each intuition 
    for fusion improve the results?
\end{enumerate}

This section first presents an overview of the proposed fusion methods 
(Section~\ref{sec:survey}) and then compares their performance 
on our data collections (Section~\ref{sec:performance}).

\begin{table*}
\centering
{\small
\caption{Summary of data-fusion methods. {\bf X} indicates
that the method considers the particular evidence.
\label{tbl:fusion}}
\begin{tabular}{|c|c|c|c|c|c|c|c|c|c|}
\hline
\multirow{2}{*}{Category} & \multirow{2}{*}{Method} & \multirow{2}{*}{\#Providers} & Source & Item  &Value& Value  & Value  & \multirow{2}{*}{Copying} \\
 &  &  & trustworthiness & trustworthiness&Popularity &  similarity &  formatting & \\

\hline
Baseline         & Vote & X & & & & & &\\
\hline
          & {\sc HUB} & X & X & & & & &\\
Web-link  & {\sc AvgLog} & X & X & & & & &\\
based     & {\sc Invest} & X & X & & & & &\\
          & {\sc PooledInvest} & X & X & & & & &\\
\hline
\multirow{3}{*}{IR based} & {\sc 2-Estimates} & X & X & & & &&\\
                           & {\sc 3-Estimates} & X & X & X & & &&\\
                           & {\sc Cosine} & X & X & & & &&\\
\hline
\multirow{4}{*}{Bayesian based} & {\sc TruthFinder} & X & X & && X & &\\
                               & {\sc AccuPr} & X & X & & & & &\\
			    & {\sc PopAccu}& X & X & &X & & &\\
                               & {\sc AccuSim} & X & X & & & X & &\\
                               & {\sc AccuFormat} & X & X & & & X & X &\\
\hline
Copying affected & {\sc AccuCopy} & X & X & & &X & X &X\\
\hline
\end{tabular}}
\end{table*}

\subsection{Review of data-fusion methods}
\label{sec:survey}
In our data collections 
each source provides at most one value on a data item and
each data item is associated with a single true value. 
We next review existing fusion methods suitable for this context.
Before we jump into descriptions of each method, we first enumerate
the many insights that have been considered in fusion.
\begin{itemize}\tightlist
  \item {\em Number of providers:} A value that is provided by
    a large number of sources is considered more likely to be true.
  \item {\em Trustworthiness of providers:} A value that is provided
    by trustworthy sources is considered more likely to be true.
  \item {\em Difficulty of data items:} The error rate on each
    particular data item is also considered in the decision.
  \item {\em Similarity of values:} The provider of a value $v$ is
    also considered as a partial provider of values similar to $v$.
  \item {\em Formatting of values:} The provider of a value $v$ is
    also considered as a partial provider of a value that subsumes $v$.
    For example, if a source typically rounds to million and provides
    ``8M'', it is also considered as a partial provider of 
    ``7,528,396''.
  \item {\em Popularity of values:} Popularity of wrong values
    is considered in the decision.
  \item {\em Copying relationships:} A copied value is ignored
    in the decision.
\end{itemize}

\eat{
\begin{table}
\centering
{\small
\caption{Notations in data fusion.
\label{tbl:notation}}
\begin{tabular}{|c|l|}
\hline
Notation          & Meaning \\
\hline
$\cal S$ & Set of sources. \\
$\cal D$ & Set of data items. \\
$V_d$ & Set of values provided for item $d$. \\
$V_s$ & Set of values provided by source $s$. \\
$D_s$ & Set of data items provided by source $s$. \\
$S_d$ & Set of providers of data item $d$. \\
$S_v$ & Set of providers of value $v$ (on data item $d$). \\
$T^k(s)$ & Trustworthiness of source $s$ in round $k$.\\
$C^k(v)$ & Vote count of value $v$ in round $k$ \\
\hline
\end{tabular}}
\end{table}}
All fusion methods more or less take a voting approach; 
that is, accumulating votes from providers for each value on the same data item
 and choosing the value with the highest vote as the true one. 
The vote count of a source is often a function of the trustworthiness
of the source. Since 
source trustworthiness is typically unknown {\em a priori}, 
they proceed in an iterative fashion: computing value vote 
and source trustworthiness in each round until
the results converge. We now briefly describe given 
a data item $d$, how each fusion method computes the vote count 
of each value $v$ on $d$ and the trustworthiness of each source $s$.
In~\cite{LDL+12b} we summarized equations applied in each method.

\smallskip
\noindent
{\sc Vote}: Voting takes the dominant
value as the true value and is the simplest strategy; 
thus, its performance is the same as the precision
of the dominant values. 
There is no need for iteration.

\smallskip
\noindent
{\sc Hub}~\cite{Kleinberg98}: Inspired by measuring web page authority
based on analysis of Web links, in {\sc HUB} the vote of a value is computed 
as the sum of the trustworthiness of its providers, while 
the trustworthiness of a source is computed as the sum of the 
votes of its provided values. 
Note that in this method the trustworthiness of a source is also affected by 
the number of its provided values.
Normalization is performed to prevent source trustworthiness and
value vote counts from growing in an unbounded manner.

\eat{
\vspace{-.15in}
{\small
\begin{eqnarray}
C^0(v)&=&.5;\\
T^k(s)&=&\sum_{v \in V_s}C^{k-1}(v); \\
C^k(v)&=&\sum_{s \in S_v}T^k(s).
\label{eqn:hub}
\end{eqnarray}
}
\vspace{-.15in}
}

\smallskip
\noindent
{\sc AvgLog}~\cite{PR10}: This method is similar to {\sc Hub} but
decreases the effect of the number of provided values 
by taking average and logarithm. Again, normalization is required.

\eat{
\vspace{-.15in}
{\small
\begin{eqnarray}
C^0(v)&=&.5;\\
T^k(s)&=&\sum_{v \in V_s}C^{k-1}(v) \cdot {\log|V_s| \over |V_s|};\\
C^k(v)&=&\sum_{s \in S_v}T^k(s).
\label{eqn:avglog}
\end{eqnarray}
}
\vspace{-.15in}
}

\smallskip
\noindent
{\sc Invest}~\cite{PR10}: A source ``invests'' its trustworthiness 
uniformly among its provided values. The vote of a value grows 
non-linearly with respect to the sum of the invested trustworthiness from its
providers. The trustworthiness of source $s$ is computed by accumulating
the vote of each provided value $v$ weighted by $s$'s contribution
among all contributions to $v$. 
 Again, normalization is required.

\eat{
\vspace{-.15in}
{\small
\begin{eqnarray}
C^0(v)&=&{|S_v| \over |S_d|};\\
T^k(s)&=&\sum_{v \in V_s}\left(C^{k-1}(v) \cdot {T^{i-1}(s)/|V_s| \over \sum_{s' \in S_v}{T^{i-1}(s')\over |V_{s'}|}}\right);\\
C^k(v)&=&\left(\sum_{s \in S_v}{T^k(s) \over |V_s|}\right)^{1.2}. 
\label{eqn:invest}
\end{eqnarray}
}
\vspace{-.15in}
}

\smallskip
\noindent
{\sc PooledInvest}~\cite{PR10}: This method is similar to 
{\sc Invest} but the vote count of each value on item $d$
is then linearly scaled such that the total vote count on $d$
equals the accumulated investment on $d$. 
With this linear scaling, normalization is not required any more.

\eat{
\vspace{-.15in}
{\small
\begin{eqnarray}
C^0(v)&=&{1 \over |V_d|}; \\
T^k(s)&=&\sum_{v \in V_s}\left(C^{k-1}(v) \cdot {T^{i-1}(s) \over |V_s| \cdot \sum_{s' \in S_v}{T^{i-1}(s')\over |V_{s'}|}}\right);\\
C_0^k(v)&=&\sum_{s \in S_v}{T^k(s) \over |V_s|}; \\
C^k(v)&=&C_0^k(v)\cdot{C_0^k(v)^{1.4} \over \sum_{v' \in V_d}{C_0^k(v')}^{1.4}}. 
\label{eqn:pooledinvest}
\end{eqnarray}
}
\vspace{-.15in}
}

\smallskip
\noindent
{\sc Cosine}~\cite{GAMS10}: This method considers the values
as a vector: for value $v$ of data item $d$, 
if source $s$ provides $v$, the corresponding position has value 1; 
if $s$ provides another value on $d$, the position has value -1;
if $s$ does not provide $d$, the position has value 0.
Similarly the vectors are defined for selected true values.
{\sc Cosine} computes the trustworthiness of a source as
the cosine similarity between the vector of its provided values
and the vector of the (probabilistically) selected values.
To improve stability, it sets the new trustworthiness as a
linear combination of the old trustworthiness and the newly computed one.

\eat{
\vspace{-.15in}
{\small
\begin{eqnarray}
C^0(v)&=&1;\\
T^k(s)&=&.8\cdot T^{k-1}(s)\\
&+&.2\cdot {\sum_{v \in V_s}C^{k-1}(v)-\sum_{d \in D_s, v \in V_d\setminus V_s}C^{k-1}(v) \over
               \sqrt{\sum_{d \in D_s}|V_d|\cdot\sum_{d \in D_s, v \in V_d}C^{k-1}(v)^2}}; \\
C^k(v)&=&{\sum_{s \in S_v}T^k(s)^3-\sum_{s \in S_d\setminus S_v}T^k(s)^3 \over
\sum_{s \in S_d}T^k(s)^3}.
\label{eqn:cosine}
\end{eqnarray}
}
\vspace{-.15in}
}

\smallskip
\noindent
{\sc 2-Estimates}~\cite{GAMS10}: {\sc 2-Estimates} also computes
source trustworthiness by aggregating value votes. It differs
from {\sc HUB} in two ways. First, if source $s$ provides value
$v$ on $d$, it considers that $s$ votes against other values on $d$
and applies a complement vote on those values. Second, it averages the vote
counts instead of summing them up. 
This method requires a complex normalization for the vote counts and trustworthiness
to the whole range of $[0,1]$.

\eat{
\vspace{-.15in}
{\small
\begin{eqnarray}
T^0(s)&=&1; \\
C^k(v)&=&{\sum_{s \in S_v}T^{k-1}(s)+\sum_{s \in S_d\setminus S_v}(1-T^{k-1}(s)) \over
                |S_d|}; \\
T^k(s)&=&{\sum_{v \in V_s}C^k(v)+\sum_{d \in D_s, v \in V_d\setminus V_s}(1-C^k(v)) \over
                \sum_{d \in D_s}|V_d|}. 
\label{eqn:2est}
\end{eqnarray}
}
\vspace{-.15in}
}

\smallskip
\noindent
{\sc 3-Estimates}~\cite{GAMS10}: {\sc 3-Estimates} improves over
{\sc 2-Estimates} by considering also {\em trustworthiness} on each value,
representing the likelihood that a vote on this
value being correct. This measure is computed iteratively together
with source trustworthiness and value vote count and similar 
normalization is applied.

\eat{
\vspace{-.15in}
{\small
\begin{eqnarray}
T^0(s)&=&1; \\
T^0(v)&=&.9; \\
C^k(v)&=&{\sum_{s \in S_v}T^{k-1}(s)T^{k-1}(v)+
                \sum_{s \in S_d\setminus S_v}(1-T^{k-1}(s)T^{k-1}(v)) \over
                |S_d|}; \\
T^k(v)&=&{\sum_{s \in S_v}{C^k(v) \over 1-T^{k-1}(s)}+
                \sum_{s \in S_d\setminus S_v}{1-C^k(v) \over 1-T^{k-1}(s)} \over |S_d|}; \\
T^k(s)&=&{\sum_{v \in V_s}{C^k(v) \over 1-T^k(v)}
                +\sum_{d \in D_s, v \in V_d\setminus V_s}{1-C^k(v) \over 1-T^k(v)} \over
                \sum_{d \in D_s}|V_d|}. 
\label{eqn:3est}
\end{eqnarray}
}
\vspace{-.15in}
}

\smallskip
\noindent
{\sc TruthFinder}~\cite{YHY08}: This method applies Bayesian analysis
and computes the probability of a value being true conditioned on
the observed providers. 
In addition, {\sc TruthFinder} considers similarity between values
and enhances the vote count of a value by those from its similar
values weighted by the similarity.

\eat{
\vspace{-.15in}
{\small
\begin{eqnarray}
T^0(s)&=&.8;\\
C^k(v)&=&\sum_{s \in S_v}-\ln(1-T^{k-1}(s)); \\ 
T^k(s)&=&\mbox{Avg}_{v \in V_s}(1-e^{-.3C^k(v)}).
\label{eqn:truthfinder}
\end{eqnarray}
}
\vspace{-.15in}

\vspace{-.15in}
{\small
\begin{eqnarray}
\hat C^k(v)&=&\sum_{v' \in V_d}C^k(v')sim(v,v').
\label{eqn:valuesim}
\end{eqnarray}
}
\vspace{-.15in}
}

\smallskip
\noindent
{\sc AccuPr}~\cite{DBS09a}: {\sc AccuPr} also applies Bayesian 
analysis. It differs from {\sc TruthFinder} in that it takes into
consideration that different values provided on the same data item
are disjoint and their probabilities should sum up to 1; 
in other words, like {\sc 2-Estimates, 3-Estimates} and {\sc Cosine},
if a source $s$ provides $v' \ne v$ on item $d$, 
$s$ is considered to vote against $v$. 
To make the Bayesian analysis possible,
it assumes that there are $N$ false values in the
domain of $d$ and they are uniformly distributed.

\eat{
\vspace{-.15in}
{\small
\begin{eqnarray}
T^0(s)&=&.8;\\
C^k(v)&=&\sum_{s \in S_v}\ln{N*T^{k-1}(S) \over 1-T^{k-1}(S)}; \\ 
T^k(s)&=&\mbox{Avg}_{v \in V_s}{e^{C^k(v)} \over \sum_{v \in V_d}e^{C^k(v)}+(N+1-|V_d|)}.
\label{eqn:truthfinder}
\end{eqnarray}
}
\vspace{-.15in}
}

\smallskip
\noindent
{\sc PopAccu}~\cite{DSS13}: {\sc PopAccu} augments {\sc AccuPr} by 
removing the assumption of having $n$ uniformly distributed false values.
It computes value distribution from the observed data.

\smallskip
\noindent
{\sc AccuSim}~\cite{DBS09a}: {\sc AccuSim} augments
{\sc AccuPr} by considering also value similarity in
the same way as {\sc TruthFinder} does.

\smallskip
\noindent
{\sc AccuFormat}: {\sc AccuFormat} augments
{\sc AccuSim} by considering also formatting of values
as we have described.

\smallskip
\noindent
{\sc AccuCopy}~\cite{DBS09a}: {\sc AccuCopy} augments
{\sc AccuFormat} by considering the copying relationships
between the sources and weighting the vote count from a source $s$
by the probability that $s$ provides the particular value independently.
In our implementation we applied the copy detection techniques in~\cite{DBS09a},
which treats sharing false values as strong evidence of copying.

\eat{
\vspace{-.15in}
{\small
\begin{eqnarray}
C^k(v)&=&\sum_{s \in S_v}I(s)\cdot\ln{N*T^{k-1}(S) \over 1-T^{k-1}(S)}. 
\label{eqn:truthfinder}
\end{eqnarray}
}
\vspace{-.15in}
}

Table~\ref{tbl:fusion} summarizes the features of different 
fusion methods. We can categorize them into five categories.
\begin{itemize}\tightlist
  \item {\em Baseline:} The basic voting strategy.
  \item {\em Web-link based:} The methods are inspired by measuring
  webpage authority based on Web links, including {\sc HUB},
  {\sc AvgLog}, {\sc Invest} and {\sc PooledInvest}.
  \item {\em IR based:} The methods are inspired by similarity
  measures in Information Retrieval, including {\sc Cosine},
  {\sc 2-Estimates} and {\sc 3-Estimates}.
  \item {\em Bayesian based:} The methods are 
  based on Bayesian analysis, including {\sc TruthFinder,
  AccuPr, PopAccu, AccuSim,} and {\sc AccuFormat}.
  \item {\em Copying affected:} The vote count computation
  discounts votes from copied values, including {\sc AccuCopy}.
\end{itemize}

Finally, note that in each method we can distinguish trustworthiness
for each attribute. For example, {\sc AccuFormatAttr} distinguishes
the trustworthiness for each attribute whereas {\sc AccuFormat} uses
an overall trustworthiness for all attributes.

\begin{table*}
\centering
{\small
\caption{Precision of data-fusion methods on one snapshot of data.
Highest precisions are in bold font and other top-3 precisions are
in bold italic font.
\label{tbl:snapshot}}
\begin{tabular}{|c|c||c|c|c|c||c|c|c|c|}
\hline
\multirow{3}{*}{Category} & \multirow{3}{*}{Method} & \multicolumn{4}{c||}{\em Stock} & \multicolumn{4}{c|}{\em Flight}  \\
\cline{3-10}
 & & prec w. & prec w/o. & Trust & Trust & prec w. & prec w/o. & Trust & Trust\\
 & & trust & trust & dev & diff & trust & trust & dev & diff  \\
\hline
Baseline & Vote & - & .908& - & - & - & .864 & -&-\\
\hline
         & {\sc HUB}&.913 & .907& .11 & .08 & .939& .857& .2& .14\\
Web-link & {\sc AvgLog}&.910& .899& .17& -.13 & .919& .839& .24& .001 \\
based & {\sc Invest} &.924& .764&.39 & -.31& .945& .754&.29 &-.12 \\
         & {\sc PooledInvest}&.924& .856&1.29 &0.29 & .945& {\bf\em .921}&17.26 &7.45 \\
\hline
         & {\sc 2-Estimates}&.910 & .903& .15 & -.14 & .87& .754& .46& -.35 \\
IR based & {\sc 3-Estimates}&.910& .905& .16& -.15 &.87& .708& .95& -.94  \\
         & {\sc Cosine}&.910 & .900& .21 & -.17 & .87& .791& .48& -.41 \\
\hline
         & {\sc TruthFinder}& .923& .911& .15 & .12 & {\bf\em.957}& .793& .25& .16 \\
         & {\sc AccuPr}&.910& .899& .14& -.11 & .91& .868& .16 & -.06 \\
	& {\sc PopAccu}&.909& .892& .14& -.11 & {\bf\em.958}& {\bf\em.925}& .17 & -.11 \\
Bayesian & {\sc AccuSim}&.918& {\bf\em .913}& .17& -.16 & .903& .844& .2& -.09 \\
based    & {\sc AccuFormat}& .918& .911& .17& -.16 & .903& .844& .2& -.09  \\
         & {\sc AccuSimAttr}&{\bf \em .950} & {\bf\em .929}& .17& -.16 &.952& .833& .19& -.08  \\
         & {\sc AccuFormatAttr}& {\bf\em .948}& {\bf .930}& .17 & -.16 &.952 & .833& .19& -.08 \\
\hline
Copying affected & {\sc AccuCopy} &{\bf.958} &.892 &.28 & -.11 & {\bf .960}& {\bf .943} & .16& -.14  \\
\hline
\end{tabular}}
\end{table*}
\subsection{Fusion performance evaluation}
\label{sec:performance}
We now evaluate the performance of various fusion methods 
on our data sets. We focus on five measures.
\begin{itemize}\tightlist
  \item {\em Precision:} The precision 
    is computed as the percentage of the output values
    that are consistent with a gold standard.
  \item {\em Recall:} The recall is computed as the percentage
    of the values in the gold standard being output as correct.
    Note that when we have fused all sources (so output
    all data items), the recall is equivalent to the precision.
  \item {\em Trustworthiness deviation:} Recall that except 
    {\sc Vote}, each method computes some trustworthiness
    measure of a source. We sampled the trustworthiness of
    each source with respect to a gold standard as it is defined in the method, and
    compared it with the trustworthiness computed by the
    method at convergence. In particular, given a source
    $s \in \cal S$, we denote by $T_{sample}(s)$ its sampled trustworthiness
    and by $T_{compute}(s)$ its computed trustworthiness,
    and compute the deviation as

\vspace{-.15in}
{\small
\begin{eqnarray}
dev({\cal S}) &=& \sqrt{{1 \over |{\cal S}|}\sum_{s \in \cal S}(T_{sample}(s)-T_{compute}(s))^2}.
\label{eqn:tolerance}
\end{eqnarray}
}
\vspace{-.15in}

  \item {\em Trustworthiness difference:} The difference
  is computed as the average computed trustworthiness for all sources
  minus the average sampled trustworthiness.
\item {\em Efficiency:} Efficiency is measured by the execution time
  on a Windows machine with Intel Core i5 processor (3.2GHz, 
4MB cache, 4.8 GT/s QPI).
\end{itemize}

\begin{figure*}[t]
\begin{minipage}[t]{0.66\textwidth}
\center
\includegraphics[width=2.2in]{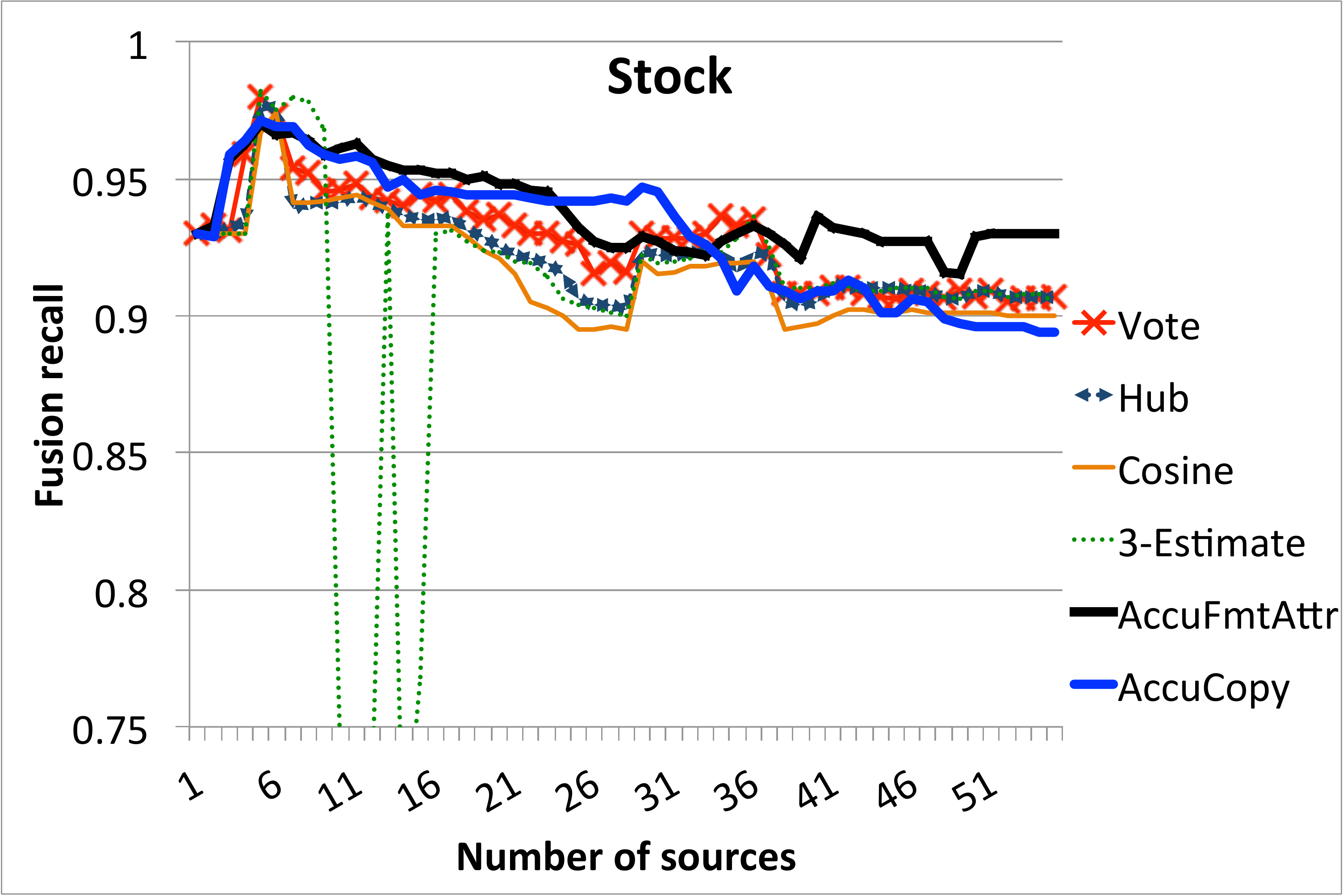}
\ \ 
\includegraphics[width=2.2in]{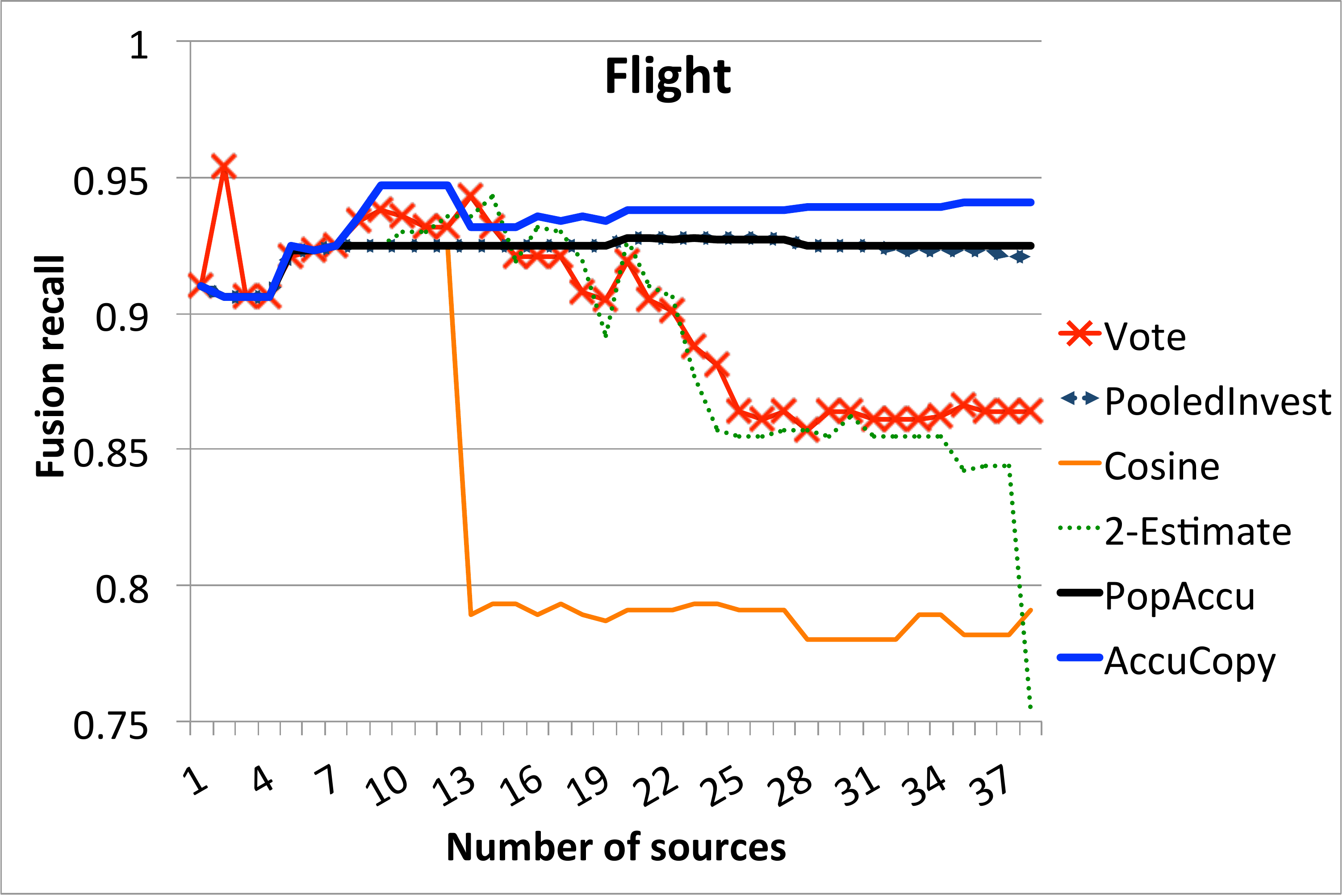}

\vspace{-.1in}
\caption{Fusion recall as sources are added.\label{fig:fusion_at_step}}
\end{minipage}
\hfill
\begin{minipage}[t]{0.33\textwidth}
\center
\includegraphics[width=2.2in]{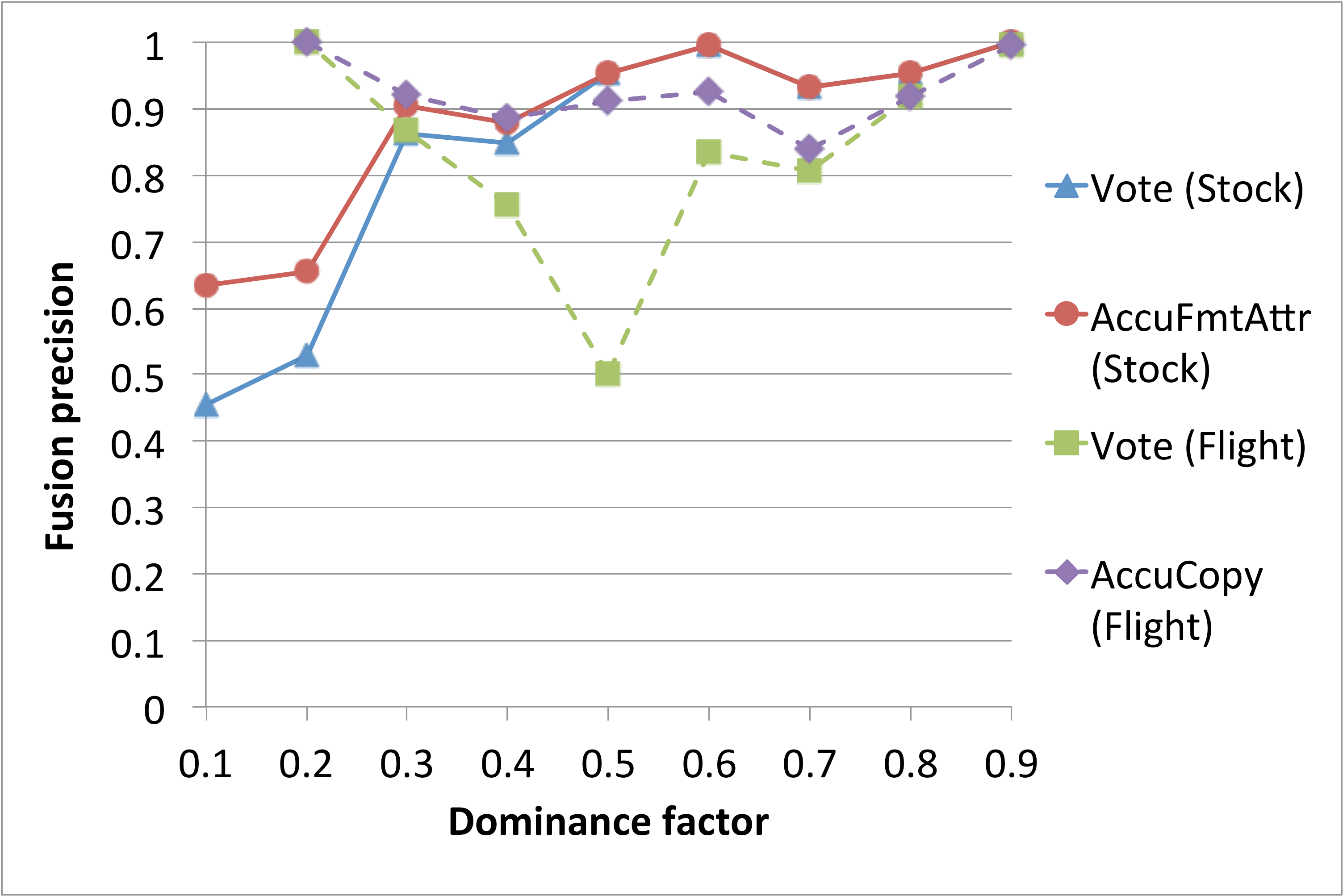}
\vspace{-.1in}
\caption{ Precision vs. dominance factor.\label{fig:precisionDom}}
\end{minipage}
\vspace{-.1in} 
\end{figure*}

\begin{table*}
\vspace{-.1in}
\centering
{\small
\caption{Comparison of fusion methods.
\label{tbl:compare}}
\begin{tabular}{|c|c||c|c|c||c|c|c|}
\hline
\multirow{2}{*}{Basic method} & \multirow{2}{*}{Advanced method} & \multicolumn{3}{c||}{\em Stock} & \multicolumn{3}{c|}{\em Flight}  \\
\cline{3-8}
 & & \#Fixed errs & \#New errs & $\Delta$Prec & \#Fixed errs & \#New errs & $\Delta$Prec \\
\hline
{\sc HUB} & {\sc AvgLog} & 3& 25& -.008& 2& 12& -.018 \\
{\sc Invest} & {\sc PooledInvest} & 376& 121& +.09& 101& 10& +.167\\
{\sc 2-Estimates} & {\sc 3-Estimates} & 6& 2& +.002& 70& 95& -.046\\
{\sc TruthFinder} & {\sc AccuSim} & 37& 32& +.002&  29& 1& +.051\\
{\sc AccuPr} & {\sc AccuSim} & 70& 31& +.014& 1& 14& -.024\\
{\sc AccuPr} & {\sc PopAccu} &7 & 26& -.007& 46& 15& +.057\\
{\sc AccuSim} & {\sc AccuSimAttr} & 47& 3& +.016& 5& 11& -.011\\
{\sc AccuSimAttr} & {\sc AccuFormatAttr} & 7& 5& +.001& 0& 0& 0 \\
{\sc AccuFormatAttr} & {\sc AccuCopy} &33 & 136&-.038 & 70& 10& +.11 \\
\hline
\end{tabular}}
\end{table*}

\eat{
}

\begin{figure}
\centering
\includegraphics[width=3.4in]{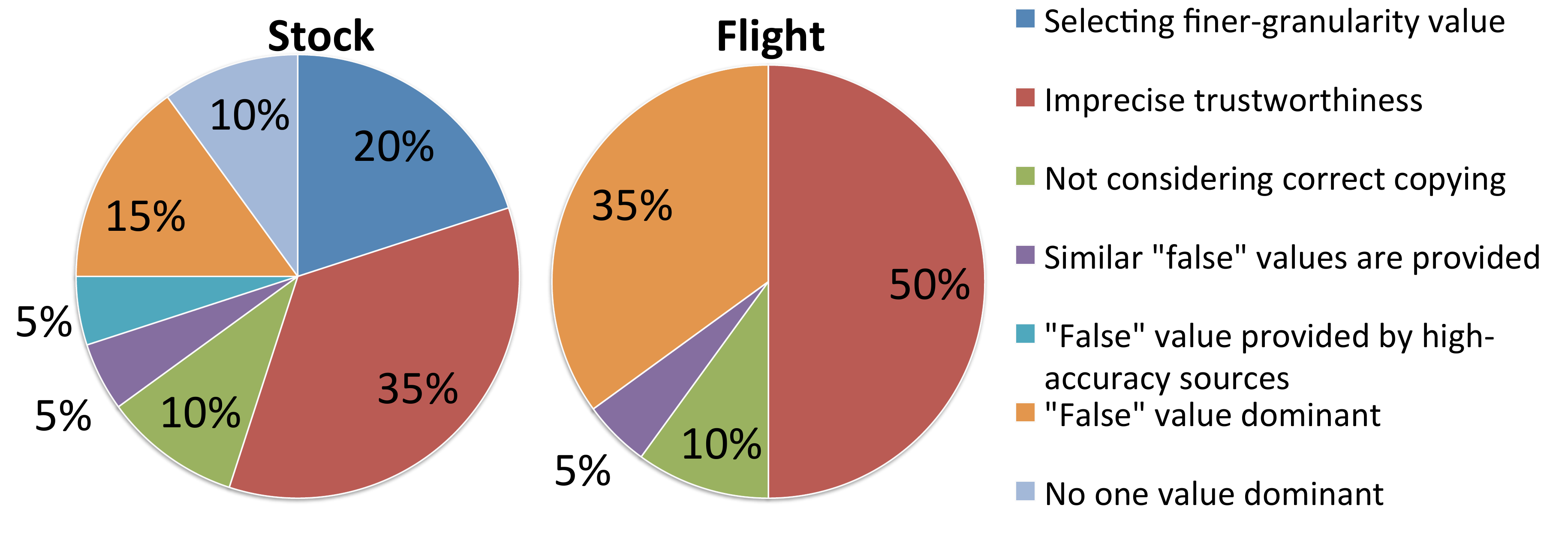}
\vspace{-.35in}
\caption{Error analysis of the best fusion method.}\label{fig:error-reasons}
\vspace{-.2in}
\end{figure}

\noindent
{\bf Precision on one snapshot:} We first consider data collected on
a particular day and use the same snapshots as in 
Section~\ref{sec:quality}.
For each data set, we computed the coverage and accuracy of
each source with respect to the gold standard
(as reported in Section~\ref{sec:quality}), and then ordered the
sources by the product of coverage and accuracy (\ie, recall).
We started with one source and gradually added sources according to
the ordering, and measured the recall. We report the following results.
First, Table~\ref{tbl:snapshot} shows the final 
precision (\ie, recall) with and without giving the sampled source 
trustworthiness as input, and the trustworthiness deviation and difference
for each method in each domain.
Second, Figure~\ref{fig:fusion_at_step} shows the recall as
we added sources on each domain; to avoid cluttering, 
for each category of fusion methods, we only plotted
for the method with the highest final recall.
Third, Table~\ref{tbl:compare} compares pairs of methods
where the second was intended to improve over the first.
The table shows for each pair how many errors by the first
method are corrected by the second and how many new
errors are introduced.
Fourth, to understand how the advanced fusion methods improve over the
baseline {\sc Vote}, Figure~\ref{fig:precisionDom}
compares the precision of {\sc Vote} and the best fusion method 
in each domain with respect to dominance factor.
Fifth, Figure~\ref{fig:error-reasons} categorizes the reasons of
mistakes for a randomly sampled 20 errors by the best fusion method
for each domain.

\smallskip
\noindent
{\em Stock data:} As shown in Table~\ref{tbl:snapshot},
for the {\em Stock} data {\sc AccuFormatAttr} obtains the
best results without input trustworthiness and it improves 
over {\sc Vote} by 2.4\% (corresponding to about 350 data items). 
As shown in Figure~\ref{fig:precisionDom},
the main improvement occurs on the data items with dominance 
factor lower than .5.
Note that on this data set the highest recall from a single
source is .93, exactly the same as that of the best fusion results.
From Figure~\ref{fig:fusion_at_step} we observe that
as sources are added, for most methods the recall
peaks at the 5th source and then gradually decreases;
also, we observe some big change for {\sc 3-Estimate}
at the 11th-16th sources.

We next compare the various fusion methods. 
For this data set, only Bayesian based methods can perform better
than {\sc Vote}; among other methods, Web-link based methods
perform worst, then {\sc AccuCopy}, then IR based methods
(Table~\ref{tbl:snapshot}). 
{\sc AccuCopy} does not perform well because it considers
copying as likely between many pairs of sources in this data set; the major
reason is that the copy-detection technique in~\cite{DBH+10a}
does not take into account value similarity, so it treats values highly similar
to the truth still as wrong and considers sharing such values as
strong evidence for copying. From Table~\ref{tbl:compare},
we observe that considering
formatting and distinguishing trustworthiness for different attributes
improve the precision on this data set, while considering
trustworthiness at the data-item level ({\sc 3-Estimate}) does not help much.

We now examine how well we estimate source trustworthiness
and the effect on fusion.
If we give the sampled source trustworthiness as input 
(so no need for iteration)
and also ignore copiers in Table~\ref{tbl:depenSources} 
when applying {\sc AccuCopy} (note that there may be other
copying that we do not know), {\sc AccuCopy} performs the best 
(Table~\ref{tbl:snapshot}).
Note that for all methods, giving the sampled trustworthiness 
improves the results. However, for most methods except
{\sc Invest, PooledInvest} and {\sc AccuCopy},
the improvement is very small; indeed, for these three methods
we observe a big trustworthiness deviation. Finally, for most methods
except {\sc Hub, PooledInvest} and {\sc TruthFinder}, 
the computed trustworthiness
is lower than the sampled one on average. This makes sense
because when we make mistakes, we compute lower trustworthiness
for most sources. {\sc TruthFinder} tends to compute very 
high accuracy, on average .97, 14\% higher than the sampled ones.

Finally, we randomly selected 20 data items on which {\sc AccuFormatAttr}
makes mistakes for examination (Figure~\ref{fig:error-reasons}). 
We found that among them, for 4 items {\sc AccuFormatAttr}
actually selects a value with finer granularity so the results cannot
be considered as wrong. Among the rest, we would be able to
fix 7 of them if we know sampled source trustworthiness, and fix 2 more
if we are given in addition the copying relationships. For the remaining
7 items, for 1 item a lot of similar false values are provided, for 1 item
the selected value is provided by high-accuracy sources,
for 3 items the selected value is provided by more than half of the sources,
and for 2 items there is no value that is dominant while the ground truth
is neither provided by more sources nor by more accurate sources than
any other value.

\smallskip
\noindent
{\em Flight data:} As shown in Table~\ref{tbl:snapshot},
on the {\em Flight} data {\sc AccuCopy}
obtains the best results without input trustworthiness
and it improves over {\sc Vote} by 9\% (corresponding to
about 550 data items, half of the mistakes made by {\sc Vote}). 
{\sc AccuCopy} does not have that many false positives for copy
detection as on {\em Stock} data because none of the attributes
here is numerical, so similar values is not a potential problem
(recall that~\cite{DBS09a} reports good results also on a
domain with non-numerical values--{\em Book}). 
As shown in Figure~\ref{fig:precisionDom}, {\sc AccuCopy} significantly
improves the precision on data items with dominance factor
in $[.4, .7)$, because it ignores copied values in fusion.
Note that on this data set the highest recall from a single
source is .91, 3.4\% lower than the best fusion results.
From Figure~\ref{fig:fusion_at_step} we observe that
as sources are added, for most methods the recall
peaks at the 9th source and then drops a lot after 
low-quality copiers are added,
but for {\sc AccuCopy}, {\sc PopAccu} and {\sc PooledInvest} the recall almost 
flattens out after the 9th source; 
also, we observe a big drop for {\sc Cosine} 
at the 14th source.

Among other methods, only {\sc PooledInvest}, {\sc PopAccu}
and {\sc AccuPr} perform better than {\sc Vote} (Table~\ref{tbl:snapshot}).
Actually, we observe that all methods perform better than
{\sc Vote} if sampled trustworthiness are given as input,
showing that the problem lies in trustworthiness computation;
this is because in this data set some groups of sources
with copying dominate the values and are considered as
accurate, while other sources that provide the true values
are then considered as less accurate. This shows that if we are
biased by low-quality copiers, considering source trustworthiness 
can bring even worse results, unless as {\sc PopAccu} does,
we keep into account the non-uniform distribution of 
false values and treat some copied values as popular false
values. Interestingly, {\sc PooledInvest}
obtains the third best results on {\em Flight} data
(but the second worst results on {\em Stock} data).
Also, we observe that considering similarity and formatting,
or distinguishing trustworthiness for each attribute does not
improve the results for this data set (Table~\ref{tbl:compare}).

If we take input trustworthiness, {\sc AccuCopy} performs 
the best (Table~\ref{tbl:snapshot}). All methods perform better with
input trustworthiness and the improvement is big.
As we have said, these are mainly because of bias from copied values.
Again, except {\sc HUB, Invest, PooledInvest} and
{\sc TruthFinder}, all other methods compute much lower trustworthiness than
the sampled ones. 

Finally, we randomly selected 20 data items on which {\sc AccuCopy}
makes mistakes for examination (Figure~\ref{fig:error-reasons}). 
We found that we would be able to
fix 10 of them if we know precise source trustworthiness, and fix 2 more
if we know correct copying relationships. For the remaining
8 items, for 1 item a lot of similar false values are provided, 
for 7 items the selected value is provided by more than half of the sources
(the value provided by the airline website is in a minority
and provided by at most 3 other sources).

\begin{figure*}[t]
\vspace{-.1in}
\begin{minipage}[b]{0.49\textwidth}
\center
\includegraphics[width=2.8in]{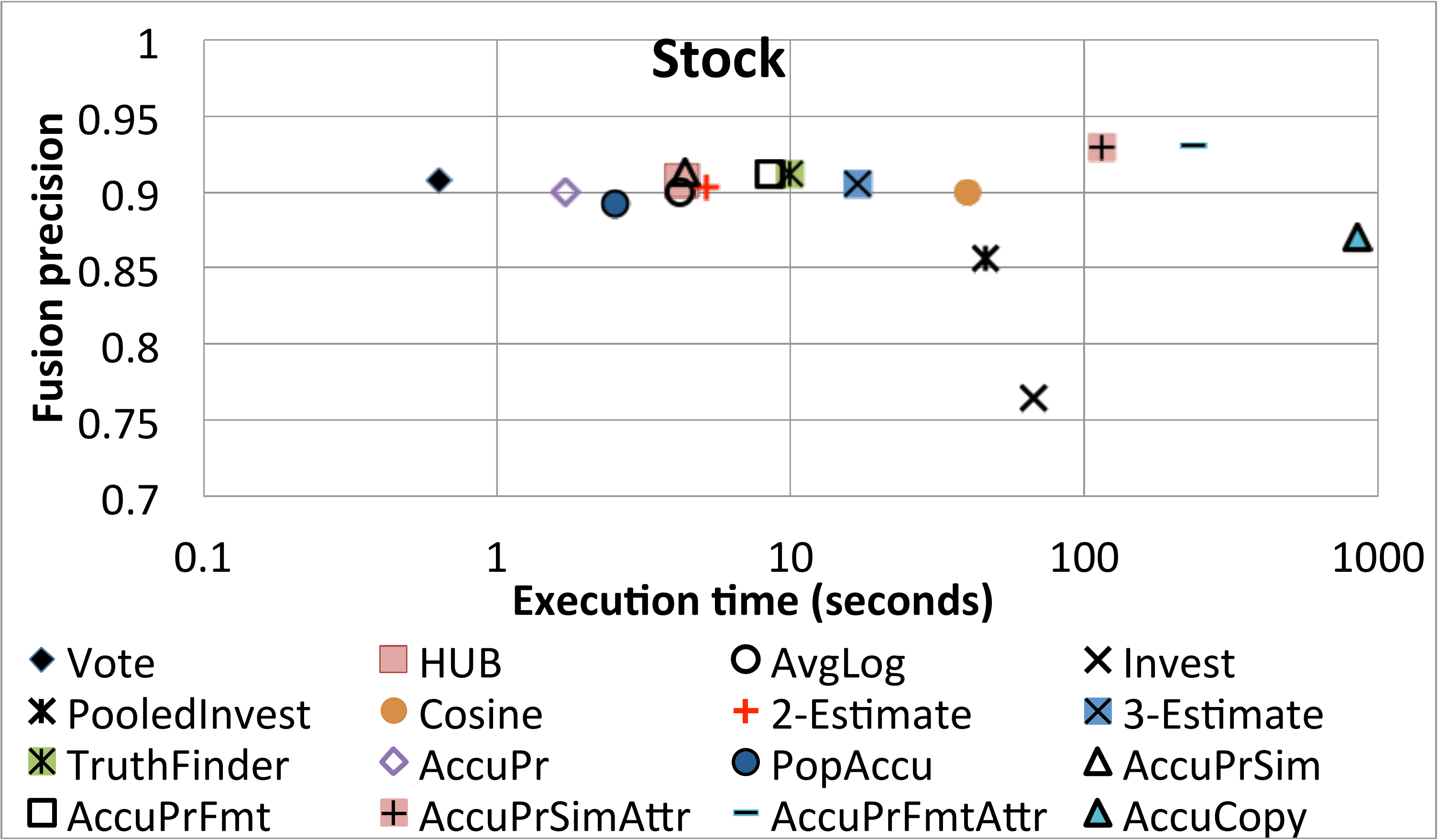}
\end{minipage}
\begin{minipage}[b]{0.49\textwidth}
\hspace{.1in}
\includegraphics[width=2.8in]{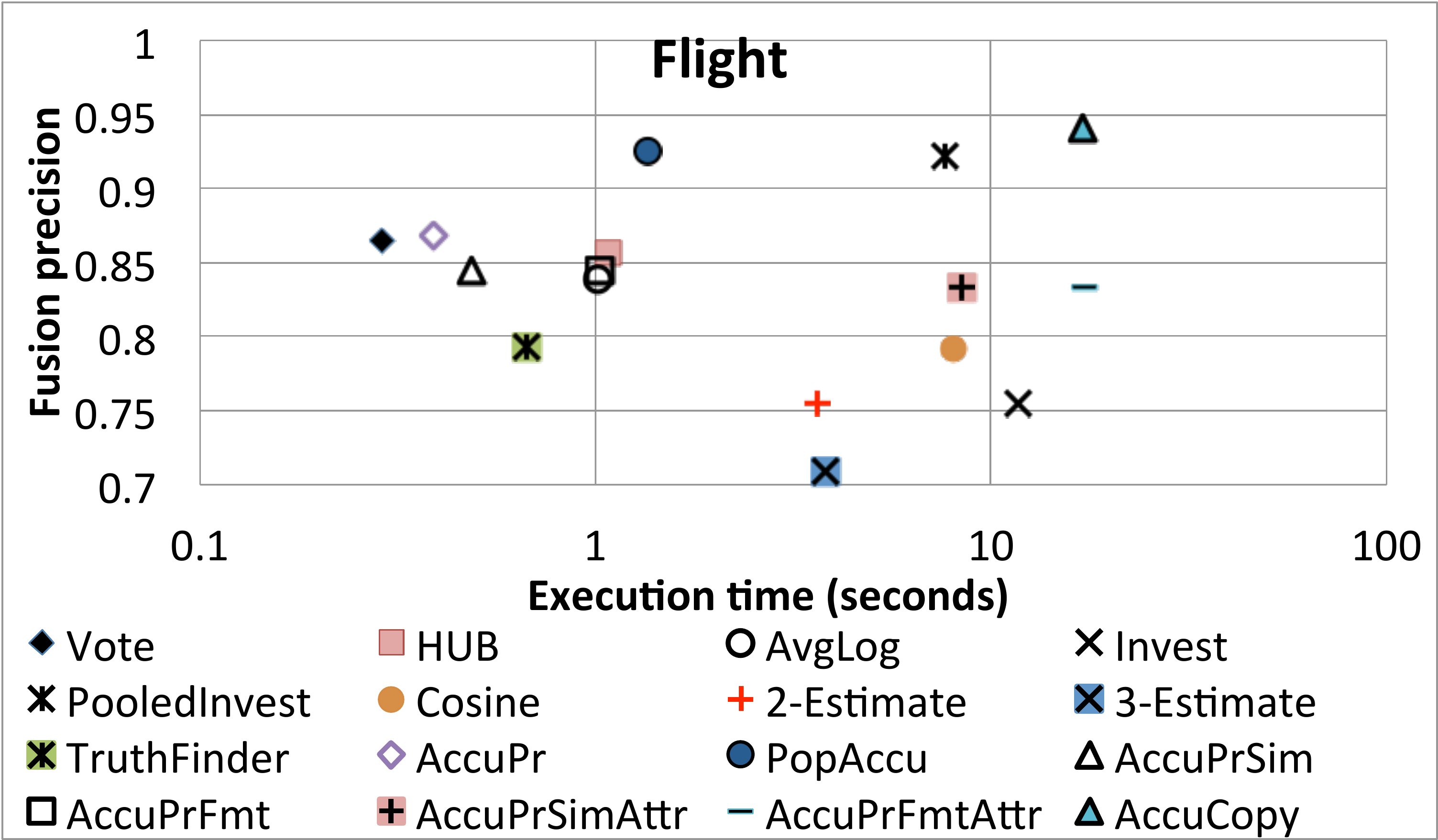}
\end{minipage}

\vspace{-.15in} 
\caption{ Fusion precision vs. efficiency.\label{fig:fusion_efficiency}}
\vspace{-.15in}
\end{figure*}

\smallskip
\noindent
{\bf Precision vs. efficiency:} Next, we examined the efficiency
of the fusion methods. Figure~\ref{fig:fusion_efficiency} plots the
efficiency and precision of each method for each domain. 

On the {\em Stock} data, {\sc Vote} finished in less than 1 second;
8 methods finished in 1-10 seconds; 4 methods, including
{\sc Invest, PooledInvest, 3-Estimate, Cosine}, finished in 10-100
seconds but did not obtain higher precision; {\sc AccuSimAttr}
and {\sc AccuFormatAttr} finished in 115 and 235 seconds respectively
while obtained the highest precision; finally, {\sc AccuCopy}
finished in 855 seconds as it in addition computes copying
probability between each pair of sources in each round,
but its precision is low.

On the {\em Flight} data, which contains fewer sources and
fewer data items than {\em Stock}, 4 methods including {\sc Vote} 
finished in less than 1 second; 9 methods finished in 
1-10 seconds; {\sc Invest} and {\sc AccuFormatAttr} finished
in 11.7 and 17.3 seconds respectively but did not obtain 
better results; {\sc AccuCopy} finished in 17 seconds and
obtained the highest precision. On this data set
{\sc AccuCopy} did not spend much longer time than {\sc AccuFormatAttr}
although it in addition computes copying probabilities, because
(1) there are fewer sources and (2) it converges in fewer rounds.

\smallskip
\noindent
{\bf Precision over time:} Finally, we ran the different fusion methods
on data sets collected on different days. Table~\ref{tbl:period}
shows for each method a summary, including average precision, 
minimum precision, and standard deviation on fusion precision over time.

Our observation in general is consistent with the results on one snapshot
of the data. {\sc AccuFormatAttr} is
the best for the {\em Stock} domain and obtains a precision of
.941 on average, whereas {\sc AccuCopy} is
the best for the {\em Flight} domain and the precision is
as high as .987.
The major difference from observations on the snapshots is 
that {\sc AccuFormatAttr} and {\sc AccuSimAttr} outperform {\sc Vote} 
on average on {\sc flight} domain. 
Finally, we observe higher deviation for {\em Flight} than
for {\em Stock}, caused by the variety of quality of copied data;
we also observe a quite high deviation for {\sc Cosine} model
on {\em Flight} data. 

\begin{table*}
\vspace{-.1in}
\centering
{\small
\caption{Precision of data-fusion methods on data over one month.
Font usage is similar to Table~\ref{tbl:snapshot}.
\label{tbl:period}}
\begin{tabular}{|c|c||c|c|c||c|c|c|}
\hline
\multirow{2}{*}{Category} & \multirow{2}{*}{Method} & \multicolumn{3}{c||}{\em Stock} & \multicolumn{3}{c|}{\em Flight} \\
\cline{3-8}
& & Avg & Min & Deviation & Avg & Min & Deviation \\
\hline
Baseline     & {\sc Vote} & .922 &.898 & .014 & .887& .861& .028 \\
\hline
                                  & {\sc HUB} &.925 &.895 & .015& .885& .850& .027\\
Web-link                          & {\sc AvgLog} &.921 &.895 & .015& .868& .838& .029\\
        based                     & {\sc Invest} &.797 & .764& .027& .786& .748& .032 \\
                                   & {\sc PooledInvest} &.871 & .831& .015& {\bf\em .979}& .921& .013\\
\hline
\multirow{3}{*}{IR based} & {\sc 2-Estimates} & .910& .811&.026 & .639& .588& .052\\
                           & {\sc 3-Estimates} & .923& .897& .014& .718& .638& .034\\
                           & {\sc Cosine} &.923 &.894 & .015& .880& .786& .086\\
\hline
                               & {\sc TruthFinder} & .930& .909& .013& .818& .777& .031 \\
                               & {\sc AccuPr} & .922& .893& .015& .893& .861& .030 \\
			& {\sc PopAccu} & .912& .884& .016& {\bf\em.972}& .779& .048 \\
Bayesian                       & {\sc AccuSim} & {\bf\em .932}& .913& .012& .866& .833& .032\\
based                          & {\sc AccuFormat} & {\bf\em .932}& .911& .012& .866& .833& .032 \\
			& {\sc AccuSimAttr} & {\bf .941}& .921& .011& .956& .833& .050 \\
			& {\sc AccuFormatAttr} & {\bf .941}& .924& .010&  .956& .833& .050\\
\hline
Copying affected & {\sc AccuCopy} & .884& .801& .036& {\bf .987} & .943& .010 \\
\hline
\end{tabular}}
\vspace{-.1in}
\end{table*}

\smallskip
\noindent
{\bf Summary and comparison:} We found that in most data sets,
the naive voting results have an even lower recall than the 
highest recall from a single source, while the best fusion method
improves over the highest source recall 
on average. We obtain very high precision for {\em Flight} (.987)
and a reasonable precision for {\em Stock} (.941).
Note however that for {\em Stock} the improvement of recall 
over a single source with the highest recall is only marginal.
Also, on all data snapshots we observe that fusing a few high-recall
sources (5 for {\em Stock}, 9 for {\em Flight})
obtains the highest recall, while adding more sources afterwards can only hurt
(reducing by 4\% for {\em Stock} and by .4\% for {\em Flight} on the snapshot).
Among the mistakes, we found that 
about 50\% can be fixed by correct knowledge of source 
trustworthiness and copying; for 10\% the selected values
have a higher granularity than the ground truth (so not erroneous);
and for the remaining 40\% we do not observe strong evidence 
from the data supporting the ground truth.

The {\em Stock} data and the {\em Flight} data represent two
types of data sets. The one represented by {\em Flight} has 
copying mainly between low-accuracy sources. On such data sets, 
considering source accuracy without copying can obtain results with 
even lower precision, while incorporating knowledge about copying can 
significantly improve the results. The data sets represented 
by {\em Stock} have copying mainly among high-accuracy sources.
In this case, ignoring copying does not
seem to hurt the fusion results much, whereas considering copying
should further improve the results; this is shown by the fact that
{\sc Vote} improves from .908 to .923 when excluding copiers, and
that {\sc AccuCopy} obtains the highest precision
(.958) among all methods when we take sampled source accuracy and discovered 
copying as input. Note however that
the low performance of {\sc AccuCopy} on {\em Stock} is because the 
copy-detection method does not handle similar values well,
so it generates lots of false positives in copy detection.
We note that other differences between the two domains
do not seem to affect the results significantly
(\eg,  despite a higher heterogeneity and more numerical values
on {\em Stock}, most methods obtain a better results on {\em Stock} data),
so we expect that our observations can generalize to other data sets.

Among the different fusion methods, we did not observe 
that one definitely dominates others on all data sets.
Similarly, for fusion-method pairs listed in Table~\ref{tbl:compare},
it is not clear that the advanced method would definitely
improve over the basic method on all data sets except for
{\sc Invest} vs. {\sc PooledInvest}, and {\sc TruthFinder} vs.
{\sc AccuSim}. For example, distinguishing trustworthiness for
different attributes helps on {\em Stock} data but not on {\em Flight} data.
However, {\sc AccuSimAttr} and {\sc AccuFormatAttr} in general
obtain higher precision than most other methods in both domains.
Typically more complex fusion methods achieve a higher fusion
precision at the expense of a (much) longer execution time.
This is affordable for off-line fusion. Certainly, 
longer execution time does not guarantee better results.

The fusion results without input trustworthiness depend 
both on how well the model performs if source
trustworthiness is given and on how well the model can estimate
source trustworthiness. In general the lower trustworthiness
deviation, the higher fusion precision, 
but there are also some exceptions.

\section{Future Research Directions}
\label{sec:extension}
Based on our observations described in 
Sections~\ref{sec:quality}-\ref{sec:fusion},
we next point out several research directions to improve
data fusion and data integration in general.

\smallskip
\noindent
{\bf Improving fusion:} 
First, considering source trustworthiness appears to be promising and can
often improve over naive voting when there is no bias from copiers.
However, we often do not know source trustworthiness {\em a priori}. 
Currently most proposed methods start from a default accuracy 
for each source and then iteratively refine the accuracy. 
However, trustworthiness
computed in this way may not be precise and it appears that 
knowing precise trustworthiness can fix nearly half of the 
mistakes in the best fusion results. Can we start with some
seed trustworthiness better than the currently employed default values 
to improve fusion results?
For example, the seed can come from sampling or based on results 
on the data items where data are fairly consistent.

Second, we observed that different fractions of data from the same
source can have different quality. The fusion results have shown
the promise of distinguishing quality of different attributes. 
On the other hand, one can imagine that data from one source may have
different quality for data items of different categories; for example,
a source may provide precise data for UA flights but low-quality
data for AA-flights. Can we automatically detect such differences
and distinguish source quality for different categories of data
for improving fusion results?

Third, we neither observed one fusion method that always dominates
the others, nor observed between a basic method and a proposed
improvement that the latter always beats the former.
Can we combine the results of different fusion models to get
better results?

Fourth, for both data sets we assumed that there is a single
true value for each data item, but in the presence of semantics ambiguity, one can
argue that for each semantics there is a true value so
there are multiple truths. Current work that considers
precision and recall of sources for fusion~\cite{ZRHG12} 
does not apply here because each source typically applies 
a single semantics for each data item. Can we effectively find all correct values
that fit at least one of the semantics and distinguish them from 
false values?

\smallskip
\noindent
{\bf Improving integration:} First, source copying not only
appears promising for improving data fusion ({\sc AccuCopy}
obtains the highest precision on {\em Flight}), but has
many other potentials to improve various aspects of data 
integration~\cite{BDD+09}. However, the copy-detection method 
proposed in~\cite{DBS09a} falls short in the presence of numerical
values as it ignores value similarity and granularity.
Can we develop more robust copy-detection methods
in such context? In addition, copy detection appears to be quite
time-consuming. Can we improve the scalability of copy detection 
for Web-scale data?

Second, even though we have tried our best to resolve heterogeneity
at the schema level and instance level manually, we still observed that
50\% of value conflicts are caused by ambiguity. In fact, observing
a lot of conflicts on an attribute from one source is a red 
flag for the correctness of schema mapping, and observing a lot of
conflicts on an object from one source is a red flag for the
correctness of instance mapping. Can we combine schema mapping,
record linkage, and data fusion to improve results of all of them?

Third, on both data sets we observed that fusion on a few high
recall sources obtains the highest recall, but on all sources 
obtains a lower recall. Such quality deterioration can also
happen because of mistakes in instance de-duplication and schema
mapping. This calls for
source selection--can we automatically select a subset of
sources that lead to the best integration results? 

\smallskip
\noindent
{\bf Improving evaluation:} No matter for data fusion,
instance de-duplication, or schema mapping, we often need
to evaluate the results of applying particular techniques. 
One major challenge in evaluation is to construct the gold standard. 
In our experiments our gold standards trust data from certain
sources, but as we observed, this sometimes puts wrong
values or coarse-grained values in the gold standard.
Can we improve gold standard construction and can we capture
our uncertainty for some data items in the gold standard? Other questions
related to improving evaluation include automatically finding and
explaining reasons for mistakes and reasons for inconsistency 
of data or schema.

\section{Conclusions}
\label{sec:conclude}
This paper is the first one that tries to understand correctness of data
in the Deep Web. We collected data in two domains where we
believed that the data should be fairly clean; to our surprise,
we observed data of quite high inconsistency and found a lot of
sources of low quality. We also applied
state-of-the-art data fusion methods to understand whether current
techniques can successfully resolve value conflicts
and find the truth. While these
methods show good potential, there is obvious space for
improvement and we suggested several promising directions for future work.

{\small
\bibliographystyle{abbrv}
\bibliography{../base}
}

\end{document}